%% file: oci.tex
\documentclass[pre,twocolumn,showpacs,superscriptaddress,preprintnumbers,floatfix]{revtex4}

\usepackage{etex}
\usepackage{ifpdf}
\usepackage{hyperref}
\usepackage{dcolumn}
\usepackage{url}
\usepackage{amsmath}
\usepackage{amscd}
\usepackage{amsfonts}
\usepackage{amssymb}
\usepackage{bm}   
\usepackage{bbm}
\usepackage{verbatim}
\usepackage{stmaryrd}
\usepackage{amsthm}
\usepackage{xcolor}
\ifpdf
  \usepackage[pdftex]{graphicx}
\else
  \usepackage[dvips]{graphicx}
\fi
\usepackage{subfigure}

\theoremstyle{plain}  \newtheorem{Lem}{Lemma}
\theoremstyle{plain}  \newtheorem*{ProLem}{Proof}
\theoremstyle{plain} 	\newtheorem{Cor}{Corollary}
\theoremstyle{plain} 	\newtheorem*{ProCor}{Proof}
\theoremstyle{plain} 	\newtheorem{The}{Theorem}
\theoremstyle{plain} 	\newtheorem*{ProThe}{Proof}
\theoremstyle{plain} 	\newtheorem{Prop}{Proposition}
\theoremstyle{plain} 	\newtheorem*{ProProp}{Proof}
\theoremstyle{plain} 	
\theoremstyle{plain}	
\theoremstyle{plain}	 
\theoremstyle{plain}	

\input{cmechabbrev}

\begin{document}

\title{Optimal Causal Inference:\\
Estimating Stored Information and Approximating Causal Architecture}

\author{Susanne Still}
\email{sstill@hawaii.edu}
\affiliation{Information and Computer Sciences,
University of Hawaii at Manoa, Honolulu, HI 96822}

\author{James P. Crutchfield}
\email{chaos@cse.ucdavis.edu}
\affiliation{Complexity Sciences Center and Physics Department,
University of California at Davis, One Shields Avenue, Davis, CA 95616}
\affiliation{Santa Fe Institute, 1399 Hyde Park Road, Santa Fe, NM 87501}

\author{Christopher J. Ellison}
\email{cellison@cse.ucdavis.edu}
\affiliation{Complexity Sciences Center and Physics Department,
University of California at Davis, One Shields Avenue, Davis, CA 95616}

\date{\today}

\begin{abstract}
We introduce an approach to inferring the causal architecture of stochastic
dynamical systems that extends rate distortion theory to use causal
shielding---a natural principle of learning. We study two distinct cases of
causal inference: optimal causal filtering and optimal causal estimation.

Filtering corresponds to the ideal case in which the probability distribution
of measurement sequences is known,
giving a principled method to approximate a system's causal structure at a
desired level of representation. We show that, in the limit in which a model
complexity constraint is relaxed, filtering finds the exact causal architecture
of a stochastic dynamical system, known as the \emph{causal-state partition}.
From this, one can estimate the amount of historical information the process
stores. More generally, causal filtering finds a graded model-complexity
hierarchy of approximations to the causal architecture. Abrupt changes in the
hierarchy, as a function of approximation, capture distinct scales of
structural organization.

For nonideal cases with finite data, we show how the correct number of
underlying causal states can be found by optimal causal estimation.
A previously derived model complexity control term allows us to correct for
the effect of statistical fluctuations in probability estimates and thereby
avoid over-fitting.

\end{abstract}


\preprint{Santa Fe Institute Working Paper 07-08-024}
\preprint{arxiv.org: 0708.1580 [cs.IT]}

\maketitle




{\bf
Natural systems compute intrinsically and produce information. This
organization, often only indirectly accessible to an observer, is reflected to
varying degrees in measured time series. Nonetheless, this information can be
used to build models of varying complexity that capture the causal architecture
of the underlying system and allow one to estimate its information processing
capabilities. We investigate two cases. The first is when a model
builder wishes to find a more compact representation than the true one. This
occurs, for example, when one is willing to incur the cost of a small increase
in error for a large reduction in model size. The second case concerns the
empirical setting in which only a finite amount of data is available. There
one wishes to avoid over-fitting a model to a particular data set.
}

\section{Introduction}

Time series modeling has a long and important history in science and
engineering. Advances in dynamical systems over the last half century led to
new methods that attempt to account for the inherent nonlinearity in many
natural phenomena \citep{Berg84,Guck83a,Wigg88a,Deva89a,Lieb93a,Ott93a,Stro94a}.
As a result, it is now well known that nonlinear systems produce highly
correlated time series that are not adequately modeled under the typical
statistical assumptions of linearity, independence, and identical distributions.
One consequence, exploited
in novel state-space reconstruction methods \citep{Pack80,Take81,Fras90b}, is
that discovering the hidden structure of such processes is key to successful
modeling and prediction \citep{Crut87a,Casd91a,Spro03a,Kant06a}. In an attempt
to unify the alternative nonlinear modeling approaches, computational mechanics
\cite{Crut88a} introduced a minimal representation---the \eM---for stochastic
dynamical systems that is an optimal predictor and from which many system
properties can be directly calculated. Building on the notion of state
introduced in Ref. \cite{Pack80}, a system's effective states are those
variables that \emph{causally shield} a system's past from its
future---capturing, in the present, information from the past that predicts the
future.

Following these lines, here we investigate the problem of learning predictive
models of time series with particular attention paid to discovering hidden
variables. We do this by using the information bottleneck method (IB)
\citep{IBN} together with a complexity control method discussed by
Ref. \citep{StillBialek2004}, which is necessary for learning from finite data.
Ref. \cite{Shal99a} lays out the relationship between computational mechanics
and the information bottleneck method. Here, we make the mathematical connection
for times series, introducing a new method.

We adapt IB to time series prediction, resulting in a method we call
\emph{optimal causal filtering}
(OCF) \footnote{A more general approach is taken in
Ref. \citep{Still09IAL}, where both predictive modeling and decision making are
considered. The scenario discussed here is a special case.}.
Since OCF, in effect, extends rate-distortion theory \citep{Shannon48} to
use causal shielding, in general it achieves an optimal balance between model
complexity and approximation accuracy. The implications of these trade-offs for
automated theory building are discussed in Ref. \citep{Still07a}. 

We show that in the important limit in which prediction is paramount and
model complexity is not restricted, OCF reconstructs the underlying process's
causal architecture, as previously defined within the framework of computational
mechanics \citep{Crut88a,Crut92c,Crut98d}. This shows that, in effect, OCF
captures a source's hidden variables and organization. The result gives
structural meaning to the inferred models. For example, one can calculate
fundamental invariants---such as, symmetries, entropy rate, and stored
information---of the original system.

To handle finite-data fluctuations, OCF is extended to \emph{optimal causal
estimation} (OCE). When probabilities are estimated from finite data, errors
due to statistical fluctuations in probability estimates must be taken
into account in order to avoid over-fitting. We demonstrate how OCF and OCI
work on a number of example stochastic processes with known, nontrivial
correlational structure.

\section{Causal States}

Assume that we are given a stochastic process $\Prob(\BiInfinity)$---a joint
distribution over a bi-infinite sequence $\BiInfinity = \Past \Future$ of
random variables. The \emph{past}, or \emph{history}, is denoted
$\Past = \ldots \MeasSymbol_{-3} \MeasSymbol_{-2} \MeasSymbol_{-1}$, while 
$\Future = \MeasSymbol_0 \MeasSymbol_1 \MeasSymbol_2 \ldots$ denotes the
\emph{future} \footnote{To save space and improve readability we use a
simplified notation that refers to infinite sequences of random variables.
The implication, however, is that one works with finite-length sequences into
the past and into the future, whose infinite-length limit is taken at
appropriate points. See, for example, Ref. \citep{Crut98d} or, for
measure-theoretic foundations, Ref. \citep{Ay05a}.}.
Here, the random variables $\MeasSymbol_t$ take on discrete values
$\meassymbol \in \ProcessAlphabet = \{ 1,2,\ldots,k\}$ and the process as a
whole is stationary.
The following assumes the reader is familiar with information theory and
the notation of Ref. \citep{Cove06a}.

Within computational mechanics, a process $\Prob(\BiInfinity)$ is viewed as a
communication channel that transmits information from the past to the future,
storing information in the present---presumably in some internal states,
variables, or degrees of freedom \cite{Crut08a}. One can ask a simple question,
then: how much information does the
past share with the future? A related and more demanding question is how we 
can infer a predictive model, given the process. Many authors have considered
such questions. Refs.  \citep{Crut01a,Crut98d,Shal99a,bialek06} review
some of the related literature.

The effective, or \emph{causal}, states $\CausalStateSet$ are determined by an
equivalence relation $\past \sim \past^\prime$ that groups all histories
together which give rise to the same prediction of the future
\citep{Crut88a,Crut98d}.
The equivalence relation partitions the space $\AllPasts$ of histories and
is specified by the set-valued function:
\begin{equation}
\epsilon(\past) =
 \{ \past^\prime: \Prob(\Future|\past) = \Prob(\Future|\past^\prime) \}
\label{CausalStateDefn}
\end{equation}
that maps from an individual history to the equivalence class
$\causalstate \in \CausalStateSet$ containing that history and all others which
lead to the same prediction $\Prob(\Future|\past)$ of the future. A causal
state $\causalstate$ includes: (i) a label $\causalstate \in \CausalStateSet$;
(ii) a set of histories
\mbox{$\Past_{\causalstate} =  \{ \past: \Prob(\Future|\past) = 
\Prob(\Future|\causalstate) \} \subset \AllPasts$}; and (iii) a future
conditional distribution $\Prob(\Future|\causalstate)$ given the state
\citep{Crut88a,Crut98d}.

Any alternative model, called a \emph{rival} $\AlternateState$,
gives a probabilistic assignment $\Prob(\AlternateState|\past)$ of histories
to its states $\alternatestate \in \AlternateStateSet$. Due to the data
processing inequality, a model can never capture more information about the
future than shared between past and future:
\begin{equation}
I[\Partition;\Future] \leq I[\Past;\Future] ~,
\label{upperbound}
\end{equation}
where $I[V,W]$ denotes the mutual information between random variables $V$ and
$W$ \citep{Cove06a}. The quantity $\EE = I[\Past;\Future]$ has been studied by
several authors and given different names, such as (in chronological order)
convergence rate of the conditional entropy \citep{Junc79}, excess entropy
\citep{Crut83a}, stored information \cite{Shaw84}, effective measure complexity
\citep{Gras86}, past-future mutual information \citep{Li91}, and predictive
information \citep{BT99}, amongst others. For a review see Ref.
\citep{Crut01a} and references therein.

The causal states $\causalstate \in \CausalStateSet$ are distinguished by the
fact that the function $\epsilon(\cdot)$ gives rise to a {\em deterministic}
assignment of histories to states:
\begin{equation}
\Prob(\causalstate|\past) = \delta_{\causalstate,\epsilon(\past)} ~,
\end{equation}
and, furthermore, by the fact that their future conditional probabilities are given by
\begin{equation}
\Prob(\Future|\causalstate) = \Prob(\Future|\past) ~,
\end{equation}
for all $\past$ such that $\epsilon(\past) = \causalstate$.
As a consequence,
the causal states, considered as a random variable $\CausalState$, capture the
full predictive information
\begin{equation}
I[\CausalState;\Future] = I[\Past;\Future] = \EE~.
\label{CS.prop.1}
\end{equation}
More to the point, they \emph{causally shield} the past and future---the past
and future are independent given the causal state:
$\Prob(\Past,\Future|\CausalState) = \Prob(\Past|\CausalState)
\Prob(\Future|\CausalState)$.

The causal-state partition has, out of all {\em equally} predictive partitions,
called {\em prescient rivals} $\PrescientState$ \cite{Crut10a}, the smallest
entropy, $\Cmu [\Partition] = H [\Partition]$:
\begin{equation}
H[\PrescientState] \geq H[\CausalState] ~,
\label{CS.prop.2}
\end{equation}
known as the \emph{statistical complexity}, $\Cmu := H[\CausalState]$. This is
amount of historical information a process stores: A process communicates $\EE$
bits from the past to the future by storing $\Cmu$ bits in the present. $\Cmu$
is one of a process's key properties; the other is its entropy rate
\citep{Cove06a}. Finally, the causal states are \emph{unique and minimal
sufficient statistics} for prediction of the time series
\citep{Crut88a,Crut98d}.

\section{Constructing Causal Models of Information Sources}
\label{OCFmotivation}

Continuing with the communication channel analogy above, models, optimal or not,
can be broadly considered to be a lossy compression of the original data. A
model captures some regularity while making some errors in describing the data.
Rate distortion theory \citep{Shannon48} gives a principled method to find a
lossy compression of an information source such that the resulting model is as
faithful as possible to the original data, quantified by a \emph{distortion
function}.
The specific form of the distortion function determines what is considered to
be ``relevant''---kept in the compressed representation---and what is
``irrelevant''---can be discarded. Since there is no universal distortion
function, it has to be assumed \emph{ad hoc} for each application. The
information bottleneck method \citep{IBN} argues for explicitly keeping the
relevant information, defined as the mutual information that the data share
with a desired relevant variable \citep{IBN}. With those choices, the
distortion function can be derived from the optimization principle, but the
relevant variable has to be specified \emph{a priori}.

In time series modeling, however, there is a natural notion of relevance: the
future data.
For stationary time series, moreover, building a model with low generalization
error is equivalent to
constructing a model that accurately predicts future data from past data.
These observations lead directly to an information-theoretic specification
for reconstructing time series models: First, introduce general model
variables $\AlternateState$ that can store, in the
present moment, the information transmitted from the past to the future.
Any set of such variables specifies a stochastic partition of $\AllPasts$
via a probabilistic assignment rule $\Prob(\AlternateState|\past)$. Second, 
require that this partition be maximally predictive. That is, it should
maximize the information $I[\AlternateState;\Future]$ that the variables
$\AlternateState$ contain about the future $\Future$. Third, the
so-constructed representation of the historical data should be a summary, i.e.,
it should not contain all of the historical information, but rather, as little
as possible while still capturing the predictive information. The information
kept about the past---$I[\Past;\AlternateState]$, the
\emph{coding rate}---measures
the model complexity or bit cost. Intuitively, one wants to find the
most predictive model at fixed complexity or, vice versa, the least complex
model at fixed prediction accuracy. These criteria are equivalent,
in effect, to causal shielding.

Writing this intuition formally reduces to the information bottleneck method,
where the relevant information is information about the future. The constrained
optimization problem one has to solve is:
\begin{equation}
\max_{\Prob(\Partition|\Past)}
  \left\{ I[\Partition;\Future] - \lambda I[\Past;\Partition] \right\} ~,
\label{OCF}
\end{equation}
where the parameter $\lambda$ controls the balance between prediction and model
complexity. The linear trade-off that $\lambda$ represents is an ad hoc
assumption \cite{Shal99a}. Its justification is greatly strengthened in the
following by the rigorous results showing it leads to the causal states and
the successful quantitative applications.

The optimization problem of Eq. (\ref{OCF}) is solved subject to the
normalization constraint:
$\sum_\AlternateState \Prob(\AlternateState|\past) = 1$, for all
$\past \in \AllPasts$. It then has a family of solutions \citep{IBN},
parametrized by the Lagrange multiplier $\lambda$, that gives the following
optimal assignments of histories $\past$ to states
$\alternatestate \in \Partition$:
\begin{equation}
\Prob_{\mathrm{opt}}(\partitionstate|\past)
 = \frac{\Prob(\partitionstate)}{Z(\past,\lambda)}
 \exp{
 	\left( -\frac{1}{\lambda}
 	\InfoGain{\Prob(\Future|\past)}{\Prob(\Future|\partitionstate)}
 	\right) ,
 }
\label{OCF_States}
\end{equation}
with
\begin{eqnarray}
\Prob(\Future|\partitionstate) & = & \frac{1}{\Prob(\partitionstate)}
 \sum_{\past \in \AllPasts} \Prob(\Future|\past)
 \Prob(\partitionstate|\past) \Prob(\past) ~\mathrm{and}\\
 \Prob(\partitionstate) & = &
 \sum_{\past \in \AllPasts} \Prob(\partitionstate|\past) \Prob(\past) ~,
\label{OCF_States_2}
\end{eqnarray}
where $\InfoGain{P}{Q}$ is the \emph{information gain} \citep{Cove06a} between
distributions $P$ and $Q$.
In the solution it plays the role of an ``energy'', effectively measuring
how different the predicted and true futures are. The more distinct, the more
information one gains about the probabilistic development of the future from
the past. That is, high energy models make predictions that deviate
substantially from the process.

These self-consistent equations are solved iteratively \citep{IBN} using a
procedure similar to the Blahut-Arimoto algorithm \citep{Arimoto72, Blahut72}.
A connection to statistical mechanics is often drawn, and the parameter
$\lambda$ is identified with a (pseudo) temperature that controls the level of
randomness; see, e.g., Ref. \citep{Rose90}. This is useful to guide intuition
and, for example, has inspired \emph{deterministic annealing}
\citep{DetermAnneal}.

We are now ready for the first observation.

\begin{Prop}
In the \emph{low-temperature regime} ($\lambda \rightarrow 0$) the
assignments of pasts to states become deterministic and are given by:
\begin{eqnarray}
\Prob_{\mathrm{opt}} (\partitionstate|\past) & = &
 \delta_{\partitionstate,\eta(\past)} ~, ~\mathrm{where}\\
 \eta(\past) & = & {\rm arg}\min_\partitionstate
 \InfoGain{\Prob(\Future|\past)}{\Prob(\Future|\partitionstate)} ~.
\label{hardassign}
\end{eqnarray}
\label{Prop:LowTempDeterministic}
\end{Prop}

\begin{ProProp}
Define the quantity
\begin{align}
D(\partitionstate) = &
  \InfoGain{\Prob(\Future|\past)}{\Prob(\Future|\partitionstate)}
  \nonumber \\
  & -  \InfoGain{\Prob(\Future|\past)}{\Prob(\Future|\eta(\past))}
  ~.
\end{align}
$D(\partitionstate)$ is positive, by definition Eq. (\ref{hardassign}) of
$\eta(\past)$. Now, write 
\begin{equation}
\Prob_{\mathrm{opt}} (\eta(\past)|\past) =
  \left(
  1 + \sum_{\partitionstate \neq \eta(\past)}
  \frac{\Prob(\partitionstate)}{\Prob(\eta(\past))}
  \exp{\left[ - \frac{D(\partitionstate)}{\lambda} \right]
  } \right)^{-1} .
\end{equation}
The sum in the r.h.s. tends to zero, as $\lambda \rightarrow 0$, assuming that
$\Prob(\eta(\past)) > 0$. Via normalization, the assignments become
deterministic.
\qed
\end{ProProp}

\section{Optimal Causal Filtering}
\label{core_results}

We now establish the procedure's
fundamental properties by connecting the solutions it determines to the
causal representations defined previously within the framework of computational mechanics. The resulting
procedure transforms the original data to a causal representation
and so we call it \emph{optimal causal filtering} (OCF).

Note first that for deterministic assignments we have
$H[\Partition|\Past] = 0$. Therefore, the information about the past becomes
$I[\Past;\Partition] = H[\Partition]$ and the objective function simplifies to 
\begin{equation}
\Fdet [\Partition] = I[\Partition;\Future] - \lambda H[\Partition] ~.
\label{OF.det}
\end{equation}

\begin{Lem}
Within the subspace of prescient rivals,
the causal-state partition maximizes $\Fdet [\widehat{\Partition}]$.
\end{Lem}

\begin{ProLem}
This follows immediately from Eqs. (\ref{CS.prop.1}) and (\ref{CS.prop.2}).
They imply that
\begin{eqnarray}
\Fdet  [\widehat{\Partition}]  &=& I[\CausalState;\Future] - \lambda H[\widehat{\Partition}] \nonumber \\
&\leq& I[\CausalState;\Future] - \lambda H [\CausalState] \nonumber \\
&=& \Fdet  [\CausalState] ~.
\end{eqnarray}
\qed
\end{ProLem}

The causal-state partition is the model with the largest value of the OCF
objective function, because it is fully predictive at minimum complexity. 
We also know from Prop. \ref{Prop:LowTempDeterministic} that in the
low-temperature limit ($\lambda \rightarrow 0$) OCF recovers a
\emph{deterministic} mapping of histories to states. We now show that this
mapping is exactly the causal-state partition of histories.

\begin{The}
OCF finds the causal-state partition of $\AllPasts$ in the low-temperature
limit, $\lambda \rightarrow 0$.
\end{The}

\begin{ProThe}
The causal-state partition, Eq. (\ref{CausalStateDefn}), always exists, and implies that there are groups of histories with
\begin{equation}
\Prob(\Future|\past) = \Prob(\Future|\epsilon(\past)) ~.
\end{equation}
We then have, for all $\past \in \Past$,
	\begin{equation}
  		\InfoGain{\Prob(\Future|\past)}{\Prob(\Future|\epsilon(\past)} = 0 ~,
	\end{equation}
and, hence,
	\begin{equation}
		\epsilon(\past) = {\rm arg}\min_\partitionstate
  		\InfoGain{\Prob(\Future|\past)}{\Prob(\Future|\partitionstate)} ~.
	\end{equation}
Therefore, we can identify $\epsilon(\past) = \eta(\past)$ in
Eq. (\ref{hardassign}), and so the assignment of histories to the causal states
is recovered by OCF:
	\begin{equation}
	\Prob_{\rm opt}(\partitionstate|\past) = \delta_{\partitionstate, \epsilon(\past)} ~.
	\end{equation}
\qed
\end{ProThe}

Note that we have not restricted the size of the set $\AlternateStateSet$ of
model states. Recall also that the causal-state partition is \emph{unique}
\citep{Crut98d}. The Lemma establishes that OCF does \emph{not} find prescient
rivals in the low-temperature limit. The prescient rivals are suboptimal in
the particular sense that they have smaller values of the objective function.
We now establish that this difference is controlled by the model size with
proportionality constant $\lambda$.

\begin{Cor}
Prescient rivals are suboptimal in OCF. The value of the objective function
evaluated for a prescient rival is smaller than that evaluated for the
causal-state model. The difference 
$\Delta \Fdet [\PrescientState] = \Fdet [\CausalState] - \Fdet[\PrescientState]$
is given by:
\begin{equation}
\Delta \Fdet [\PrescientState] =
  \lambda \left( \Cmu [\PrescientState] - \Cmu [\CausalState] \right) \geq 0 ~.
\end{equation}
\end{Cor}

\begin{ProCor}
\begin{align}
\Delta \Fdet [\PrescientState] &= \Fdet [\CausalState] - \Fdet [\PrescientState]  \\
  &= I[\CausalState;\Future] -  I[\PrescientState;\Future]
	- \lambda H[\CausalState] + \lambda H[\PrescientState] \\
  &= \lambda \left( \Cmu [\PrescientState] - \Cmu [\CausalState] \right) ~.
\end{align}
Moreover, Eq. (\ref{CS.prop.2}) implies that $\Delta \Fdet \geq 0$.
\qed
\end{ProCor}

So, we see that for $\lambda = 0$, causal states and all other prescient rival
partitions are degenerate. This is to be expected as at $\lambda = 0$ the
model-complexity constraint disappears. Importantly, this means that maximizing
the predictive information alone, without the appropriate constraint on model
complexity does not suffice to recover the causal-state partition. 

\section{Examples}
\label{examples}

We study how OCF works on a series of example stochastic processes of 
increasing statistical sophistication. We compute the optimal solutions and
visualize the trade-off between predictive power and complexity of the model
by tracing out a curve similar to a rate-distortion curve
\citep{Arimoto72, Blahut72}: For each value of $\lambda$, we evaluate both the
model's coding rate $I[\Past;\AlternateState]$ and its predicted information
$I[\AlternateState;\Future]$ at the optimal solution and plot them against each
other. The resulting curve in the \emph{information plane} \citep{IBN}
separates the
feasible from the infeasible region: It is possible to find a model that is
more complex at the same prediction error, but not possible to
find a less complex model than that given by the optimum. In analogy to a
rate-distortion curve, we can read off the maximum amount of information about
the future that can be captured with a model of fixed complexity. Or,
conversely, we can read off the smallest representation at fixed
predictive power.

The examples in this and the following sections are calculated by solving the
self-consistent Eqs. (\ref{OCF_States}) to (\ref{OCF_States_2})
iteratively \footnote{The algorithm follows that used in the information
bottleneck \citep{IBN}. The convergence arguments there apply to the OCF
algorithm.}
at each value of $\lambda$. To trace out the curves, a deterministic annealing
\citep{DetermAnneal} scheme is implemented, lowering $\lambda$ by a fixed
annealing rate. Smaller rates cost more computational time, but allow one to
compute the rate-distortion curve in greater detail, while larger rates result
in a rate-distortion curve that gets evaluated in fewer places and hence looks
coarser. In examples, naturally, one can only work with finite length past and
future sequences: $\finpast{K}$ and $\finfuture{L}$, where $K$ and $L$ give
their lengths, respectively.

\subsection{Periodic limit cycle: A predictable process}

\begin{figure*}
\begin{center}
\resizebox{!}{2.50in}{\includegraphics{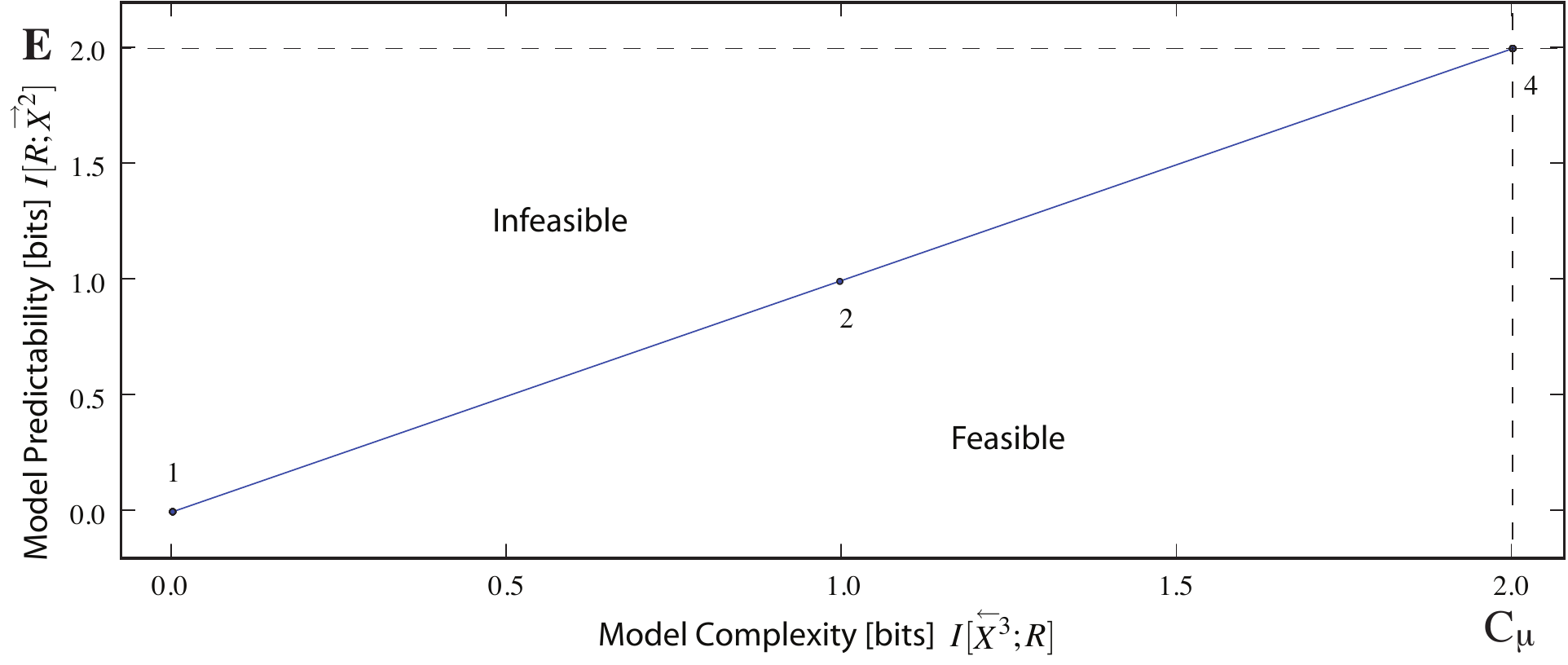}}
\end{center}
\caption{Model predictability $I[\AlternateState;\FinFuture{L}]$ versus
  model complexity (size)
  $I[\FinPast{K};\AlternateState]$ trade-off under OCF for the exactly
  predictable period-$4$ process: $(0011)^{\infty}$. Monitored in the
  information plane. The horizontal dashed line is the full predictive
  information ($\EE = I[\FinPast{3};\FinFuture{2}] = 2$ bits) and the vertical
  dashed line is the block entropy ($H[\FinPast{3}] = 2$ bits),
  which is also the statistical complexity $\Cmu$.
  The data points represent solutions at various $\lambda$.
  Lines connect them to help guide the eye only.
  Histories of length $K = 3$ were used, along with futures of length $L = 2$.
  In this and the following information plane plots, the integer labels
  $N_c$ indicate the first point at which the effective number of
  states used by the model equals $N_c$.
  }
\label{fig:Period4.MI}
\end{figure*}

\begin{figure*}
\begin{center}
\resizebox{!}{2.50in}{\includegraphics{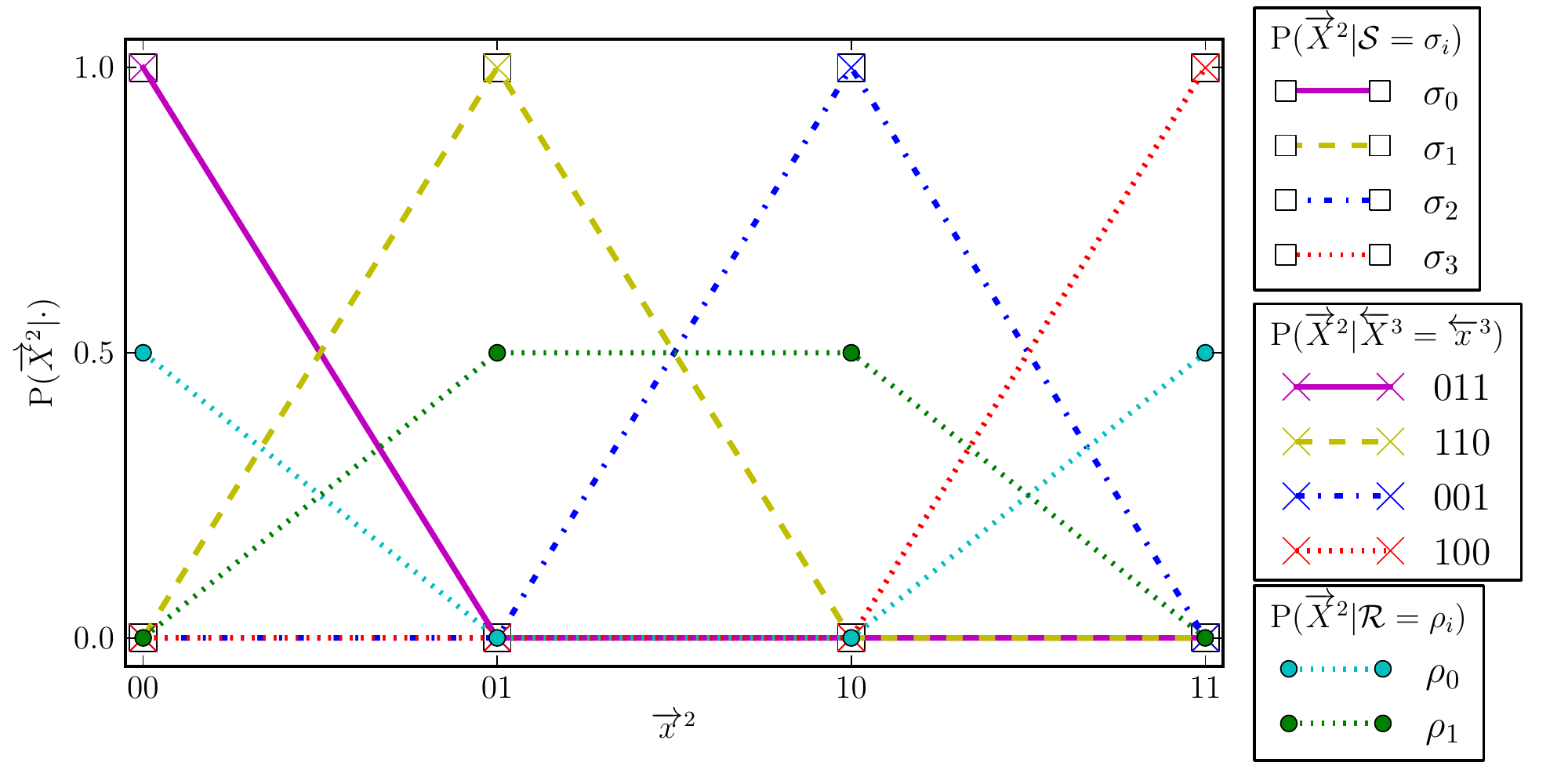}}
\end{center}
\caption{Morphs $\Prob(\FinFuture{2}|\cdot)$ for the period-$4$ process: The
  $2$-state approximation (circles) compared to the $\delta$-function morphs
  for the $4$ causal states (boxes).
  The morphs $\Prob(\FinFuture{2}|\causalstate)$ for the two-state approximation
  are $(1/2,0,0,1/2)$ and $(0,1/2,1/2,0)$ and for the four-state
  case $(1,0,0,0)$, $(0,1,0,0)$, $(0,0,1,0)$, and $(0,0,0,1)$.
  Histories of length $K = 3$ were used, along with futures of length
  $L = 2$ (crosses).
  }
\label{fig:Period4.morphs}
\end{figure*}

We start with an example of an exactly periodic process, a limit cycle
oscillation. It falls in the class of deterministic and time reversible
processes, for which the rate-distortion curve can be computed
analytically---it lies on the diagonal \citep{Still07a}.
We demonstrate this with a numerical example. Figure \ref{fig:Period4.MI}
shows how OCF works on a period-four process: $(0011)^{\infty}$.
(See Figs. \ref{fig:Period4.MI} and \ref{fig:Period4.morphs}.) There are
exactly two bits of predictive information $I[\Past;\Future]$ to be
captured about future words of length two
(dotted horizontal line). This information describes the phase of the
period-four cycle. To capture those two bits, one needs exactly four
underlying causal states and a model complexity of $\Cmu = 2$ bits
(dotted vertical line). 

The curve is the analog of a rate-distortion curve, except that the
information plane swaps the horizontal and vertical axes---the coding rate
and distortion axes. (See Ref.  \citep{Still07a} for the direct use of the
rate-distortion curve.) The value of $I[\AlternateState;\FinFuture{2}]$
(the ``distortion''), evaluated at the optimal distribution,
Eq. (\ref{OCF_States}), is plotted versus $I[\FinPast{3};\AlternateState]$
(the ``code rate''),
also evaluated at the optimum. Those are plotted for different values of
$\lambda$ and, to trace out the curve, deterministic annealing is implemented.
At large $\lambda$, we are in the lower left of the curve---the compression is
extreme, but no predictive information is captured. A single state model, a
fair coin, is found as expected. As $\lambda$ decreases (moving to the right),
the next distinct point on the curve is for a two-state model, which discards
half of the information. This comes exactly at the cost of one predictive bit.
Finally, OCF finds a four-state model that captures all of the predictive
information at no compression. The numbers next to the curve indicate the
first time that the effective number of states increases to that value. 

The four-state model captures the two bits of predictive information. But
compressed to one bit (using two states), one can only capture one bit of
predictive information. The information curve falls onto the diagonal---a
straight line that is the worst case for possible beneficial trade-offs
between prediction error and model complexity \citep{Still07a}. 

In Fig. \ref{fig:Period4.morphs}, we show the best two-state model compared to
the full (exact) four-state model. One of the future conditional probabilities
captures zero probability events of ``odd'' $\{01,10\}$  words, assigning equal
probability to the ``even'' $\{00,11\}$ words. The other one captures zero
probability events of even words, assigning equal probability to the odd words.
This captures the fundamental determinism of the process: an odd word never
follows an even word and vice versa. The overall result illustrates how the
actual long-range correlation in the completely predictable period-$4$ 
sequence is represented by a smaller \emph{stochastic} model. While in the
four-state model the future conditional probabilities are $\delta$-functions,
in the two-state approximate model they are mixtures of those
$\delta$-functions. In this way, OCF converts structure to randomness 
when approximating underlying states with a compressed model; cf.
the analogous trade-off discussed in Ref. \citep{Crut01a}.

\begin{figure*}[ht]
\centering
\resizebox{!}{2.50in}{\includegraphics{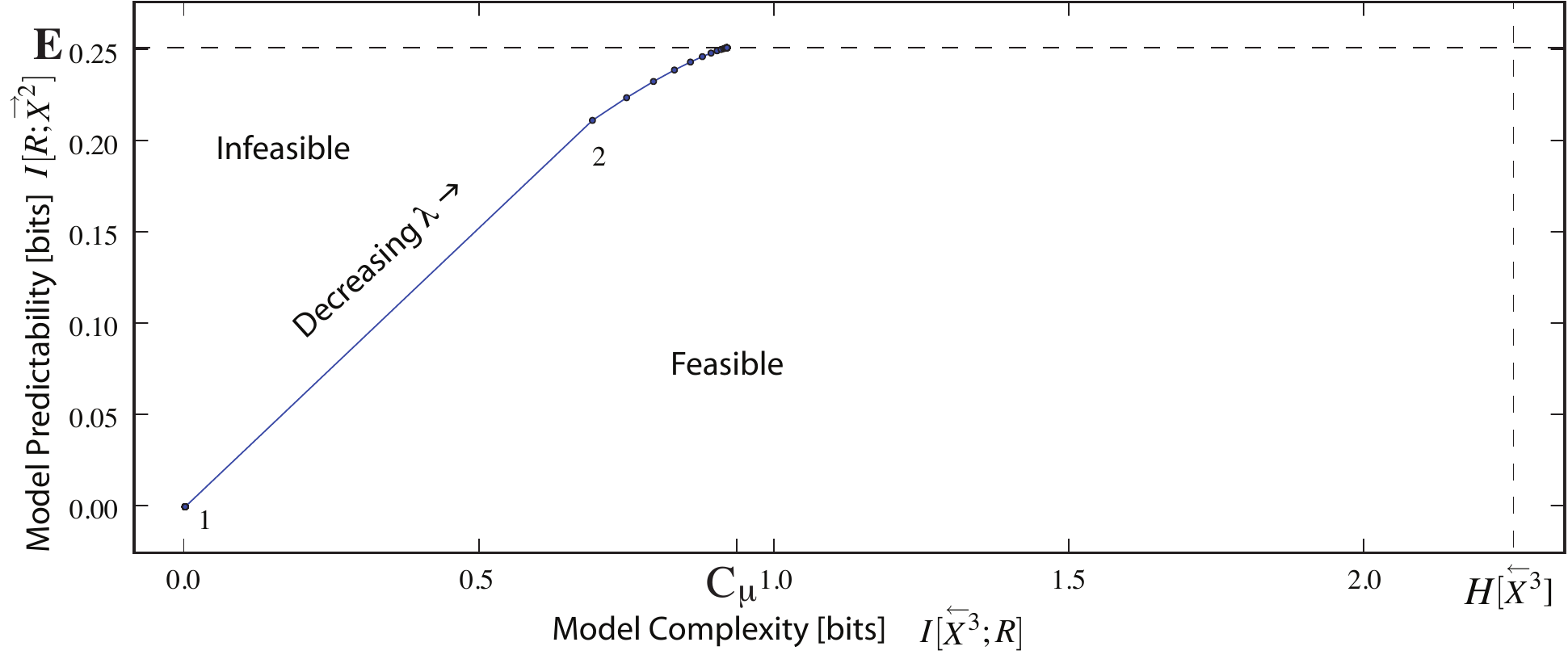}}
\caption{OCF's behavior monitored in the information
  plane---$I[\AlternateState;\FinFuture{2}]$ versus
  $I[\FinPast{3};\AlternateState]$---for the Golden Mean Process.
  The correct two-state model is found.
  Histories of length $K = 3$ were used, along with futures of
  length $L = 2$. The
  horizontal dashed line is the full predictive information
  $\EE \approx I[\FinPast{3};\FinFuture{2}] = I[\CausalState;\FinFuture{2}] \approx 0.25$
  bits which, as seen, is an upper bound on 
  $I[\AlternateState;\FinFuture{2}]$.
  The exact value is $\EE = I[\Past;\Future] = 0.2516$ bits \cite{Crut08b}.
  Similarly, the vertical dashed line is
  the block entropy $H[\FinPast{3}] \approx 2.25$ bits which is an upper
  bound on the retrodictive information $I[\FinPast{3};\AlternateState]$.
  The statistical complexity $\Cmu \approx 0.92$ bits, also an
  upper bound, is labeled.
  The annealing rate was $0.952$.
  }
\label{fig:OCFGMPInfoPlane}
\end{figure*}

\begin{figure*}[ht]
\centering
\resizebox{!}{2.50in}{\includegraphics{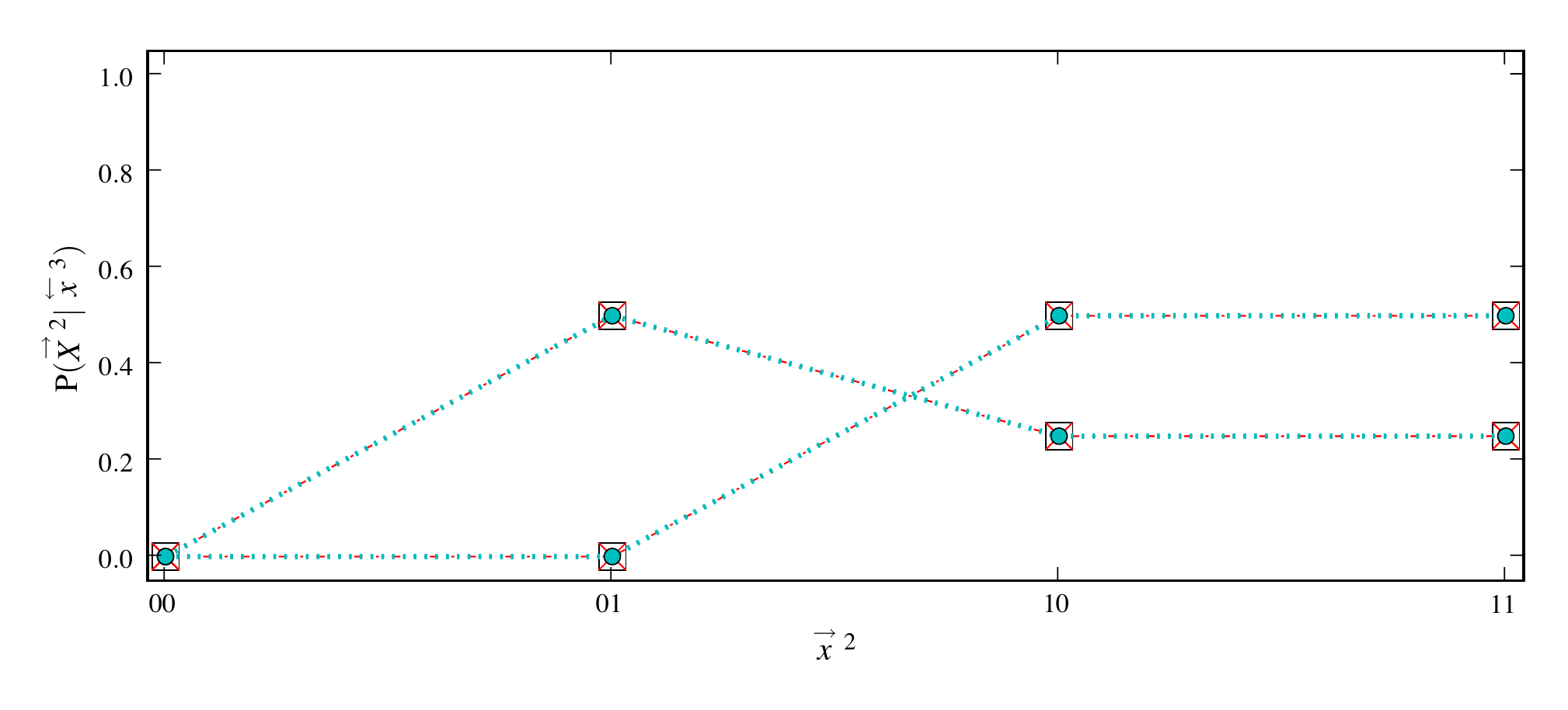}}
\caption{Future conditional probabilities $\Prob(\FinFuture{2}|\cdot)$
  conditioned on causal states $\causalstate \in \CausalStateSet$ (boxes) and
  on the OCF reconstructed states $\alternatestate \in \AlternateStateSet$
  (circles) for the Golden Mean Process. As an input to OCF, future conditional
  probabilities $P(\FinFuture{2}|\finpast{3})$ calculated from histories of
  length $K = 3$ were used (crosses).}
\label{fig:OCFGMPMorphs}
\end{figure*}

\subsection{Golden Mean Process: A Markov chain}

The Golden Mean (GM) Process is a Markov chain of order one. As an information
source, it produces all binary strings
with the restriction that there are never consecutive $0$s. The GM Process
generates $0$s and $1$s with equal probability, except that once a $0$ is
generated, a $1$ is always generated next. One can write down a simple
two-state Markov chain for this process; see, e.g., Ref. \citep{Crut01a}.

Figures \ref{fig:OCFGMPInfoPlane} and \ref{fig:OCFGMPMorphs}
demonstrate how OCF reconstructs the states of the GM process.
Figure \ref{fig:OCFGMPInfoPlane} shows the behavior of OCF in the
information plane. At very high temperature
($\lambda \rightarrow \infty$, lower left corner of the curve) compression
dominates over prediction and the resulting model is most compact, with only
one effective causal state. However, it contains no information
about the future and so is a poor predictor. As $\lambda$
decreases (moving right), OCF
reconstructs increasingly more predictive and more complex models. The curve
shows that the information about the future, contained in the optimal
partition, increases (along the vertical axis) as the model increases in
complexity (along the horizontal axis). There is a transition to two effective
states: the number $2$ along the curve denotes the first occurrence
of this increase.
As $\lambda \rightarrow 0$, prediction comes to dominate and OCF finds a fully
predictive model, albeit one with the minimal statistical complexity, out of
all possible state partitions that would retain the full predictive information.
The model's complexity---$\Cmu \approx 0.92$ bits---is 41\% of the maximum,
which is given by the entropy of all possible pasts of length $3$:
$H[\FinPast{3}] \approx 2.25$ bits. The remainder (59\%) of the information is
nonpredictive and has been filtered out by
OCF. Figure \ref{fig:OCFGMPMorphs} shows the future conditional probabilities,
associated with the partition found by OCF, as $\lambda \rightarrow 0$,
corresponding to $\Prob( \FinFuture{2} | \alternatestate )$ (circles). These
future conditional probabilities overlap with the true (but not known to the
algorithm) causal-state future conditional probabilities
$\Prob(\FinFuture{2}|\causalstate)$ (boxes) and so 
demonstrate that OCF finds the causal-state partition. 

\subsection{Even Process: A hidden Markov chain}
\label{sec:EvenProcess}

Now, consider a hidden Markov process: the {\em Even Process} \citep{Crut01a},
which is a stochastic process whose support (the set of allowed sequences)
is a symbolic dynamical system called the \emph{Even system}.
The Even system generates all binary strings consisting
of blocks of an even number of $1$s bounded by $0$s. Having observed a
process's sequences, we say that a word (finite sequence of symbols)
is \emph{forbidden} if it never occurs.  A word is an \emph{irreducible
forbidden word} if it contains no proper subwords which are themselves
forbidden words.  A system is \emph{sofic} if its list of irreducible
forbidden words is infinite.  The Even system is one such sofic system, since
its set $\mathcal{F}$ of irreducible forbidden words is infinite:
$\mathcal{F} = \{ 01^{2n+1}0, n = 0 , 1, \ldots \}$. Note that no finite-order
Markovian source can generate this or, for that matter, any other strictly
sofic system \citep{Crut01a}. The Even Process then associates probabilities
with each of the Even system's sequences by choosing a $0$ or $1$ with
fair probability after generating either a $0$ or a pair of $1$s.  The
result is a \emph{measure sofic process}---a distribution over a
sofic system's sequences.

\begin{figure*}[ht]
\centering
\resizebox{!}{2.50in}{\includegraphics{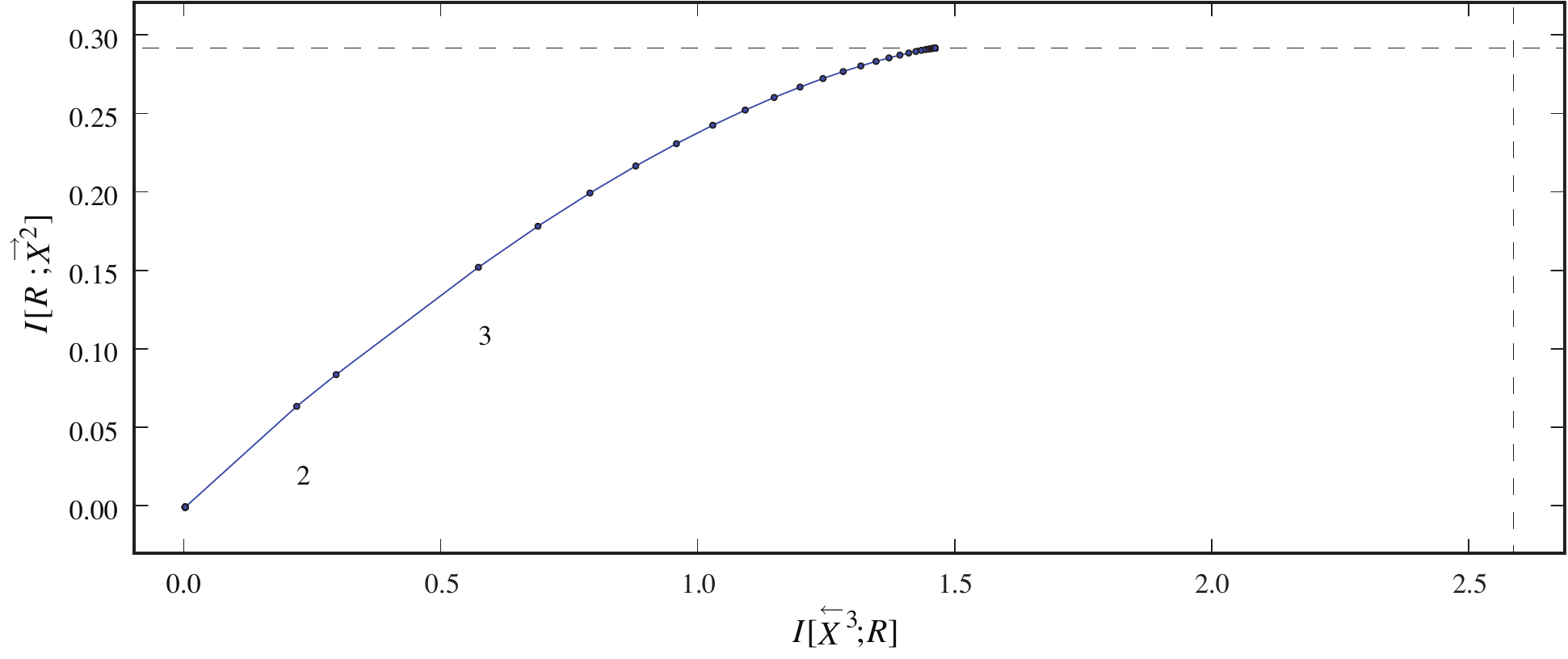}}
\caption{OCF's behavior inferring the Even Process: monitored in the
  information plane---$I[\AlternateState;\FinFuture{2}]$ versus
  \mbox{$I[\FinPast{3};\AlternateState]$}.
  Histories of length $K = 3$ were used, along with futures of
  length $L = 2$. The horizontal dashed line is the full predictive information
  $I[\FinPast{3};\FinFuture{2}] \approx 0.292$ bits which, as seen, is an
  upper bound on the estimates $I[\AlternateState;\FinFuture{2}]$. Similarly,
  the vertical dashed line is the block entropy $H[\FinPast{3}] \approx 2.585$
  bits which is an upper bound on the retrodictive information
  $I[\FinPast{3};\AlternateState]$.
  }
\label{fig:OCFExampleInfoPlane}
\end{figure*}

\begin{figure*}[ht]
\centering
\resizebox{!}{2.50in}{\includegraphics{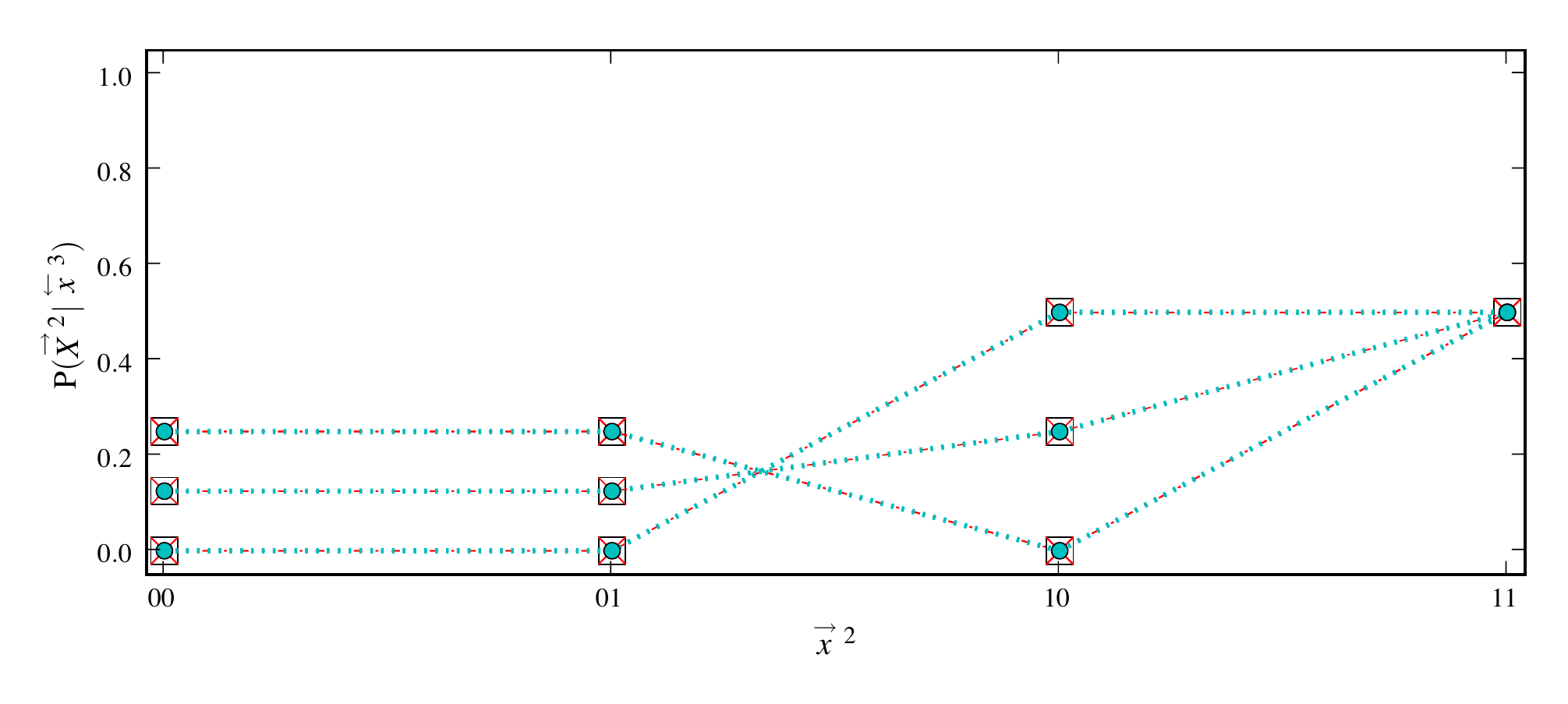}}
\caption{Future future conditional probabilities $\Prob(\FinFuture{2}|\cdot)$
  conditioned on causal states $\causalstate \in \CausalState$ (boxes) and on
  the OCF-reconstructed states $\alternatestate \in \AlternateState$ (circles)
  for the Even Process. As an input to OCF, future conditional probabilities
  $P(\FinFuture{2}|\finpast{3})$ calculated from histories of length $K = 3$
  were used (crosses).}
\label{fig:OCFExampleMorphs}
\end{figure*}

As in the previous example, for large $\lambda$, OCF applied to the Even
Process recovers a small, one-state model with poor predictive quality;
see Fig. \ref{fig:OCFExampleInfoPlane}. As $\lambda$ decreases there are 
transitions to larger models that capture increasingly more information about
the future. (The numbers along the curve again indicate the points
of first transition to more
states.) With a three-state model OCF captures the full predictive information
at a model size of 56\% of the maximum. This model is exactly the causal-state
partition, as can be seen in Fig. \ref{fig:OCFExampleMorphs} by comparing the
future conditional probabilities of the OCF model (circles) to the true
underlying causal states (boxes), which are not known to the algorithm.

The correct \eM\ model of the Even Process has four causal states: two transient
and two recurrent. At the finite past and future lengths used here, OCF picks up
only one of the transient states and the two recurrent states. It also assigns
probability to all three. This increases the effective state entropy
($H[\AlternateState] \approx 1.48$ bits) above the statistical complexity
($\Cmu = 0.92$ bits) which is only a function of the two recurrent states,
since asymptotically ($K \rightarrow \infty$) the transient states have zero
probability.

There is an important lesson in this example for general time-series modeling,
not just OCF. Correct inference of even finite-state, but measure-sofic
processes requires using hidden Markov models. Related consequences of this,
and one resolution, are discussed at some length for estimating ``nonhidden''
Markov models of sofic processes in Ref. \cite{Stre07a}.

\subsection{Random Random XOR: A structurally complex process}

The previous examples demonstrated our main theoretical result: In the
limit in which it becomes crucial to make the prediction error very small,
at the expense of the model size, the OCF algorithm captures all of the
structure inherent in the process by recovering the causal-state partition.

However, if we allow (or prefer) a model with some finite prediction error,
then we can make the model substantially smaller. We have already seen what
happens in the worst case scenario, for a periodic process. There, each
predictive bit costs exactly one bit in terms of model size. However, for
highly structured processes, there exist situations in which one can compress
the model substantially at essentially no loss in terms of predictive power.
(This is called \emph{causal compressibility} \citep{Still07a}.) The Even
Process is an example of such an information source: The statistical complexity
$H[\CausalState]$ of the causal-state partition is smaller than the total
available historical information---the entropy of the past
$H[\FinPast{K}]$.

Now, we study a process that requires keeping \emph{all} of the historical
information to be maximally predictive, which is the same as stating
$\Cmu(\AlternateState) = H[\FinPast{K}]$. (Precisely, we mean given the finite
past and future
lengths we use.) Nonetheless, there is a systematic ordering
of models of different size and different predictive power given by the
rate-distortion curve, as we change the parameter $\lambda$ that controls how
much of the future fluctuations the model considers to be random; i.e., which
fluctuations are considered indistinguishable. Naturally, the trade-off, and
therefore the shape of the rate-distortion curve, depends on and reflects the
source's organization.

\begin{figure*}
\begin{center}
\resizebox{!}{2.50in}{\includegraphics{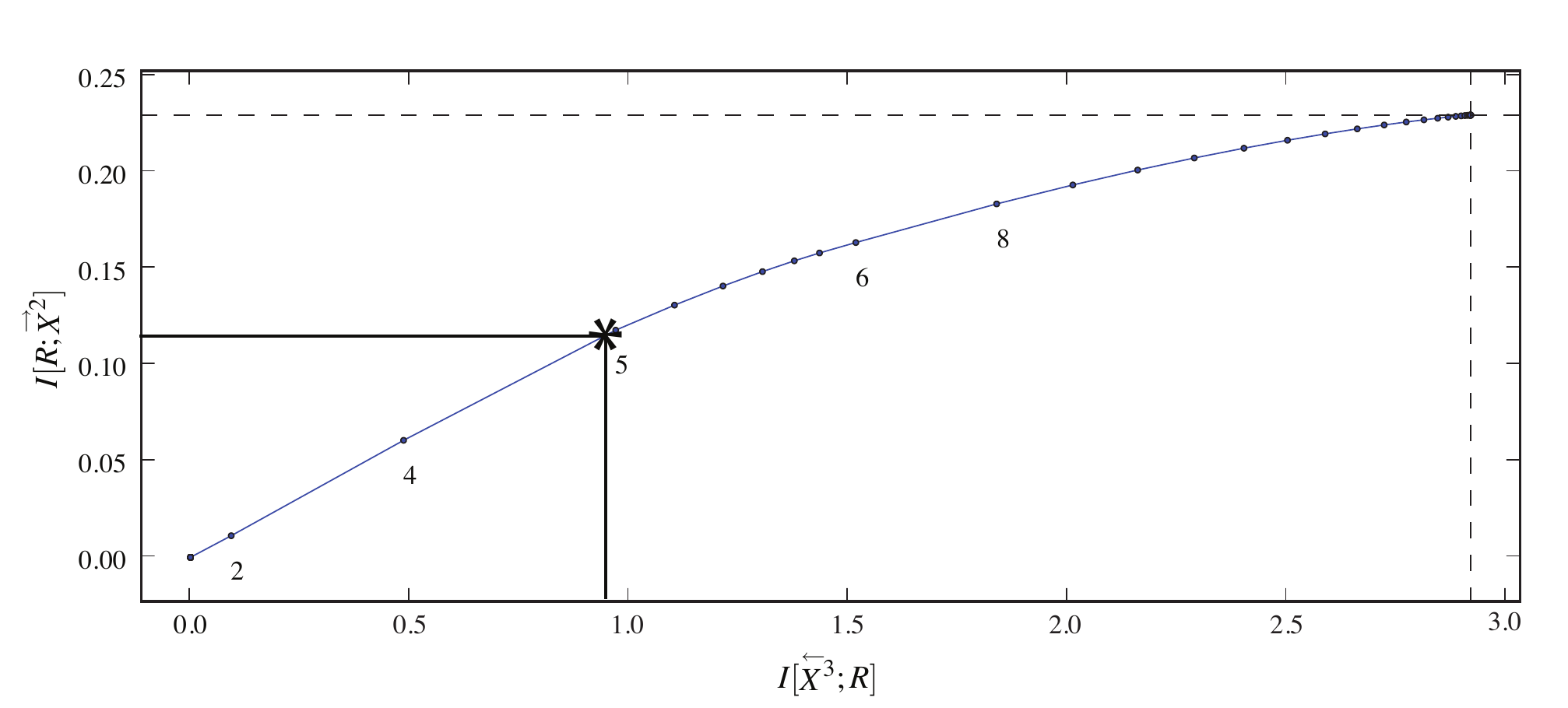}}
\end{center}
\caption{Prediction versus structure trade-off under OCF for the random-random
  XOR (RRXOR) process, as monitored in the information plane. As above, the
  horizontal dashed line is the predictive information ($\approx 0.230$ bits)
  and the vertical dashed line is the block entropy ($\approx 2.981$ bits).
  Histories of length $K = 3$ were used, along with futures of length $L = 2$.
  The asterisk and lines correspond to the text: they serve to show how the
  predictive power and the complexity of the best four state model, the future
  conditional probabilities of which are depicted in
  Fig. \ref{fig:RRXORMorphs4}.
  }
\label{fig:RRXORMInfoPlane}
\end{figure*}

\begin{figure*}
\begin{center}
\resizebox{!}{2.50in}{\includegraphics{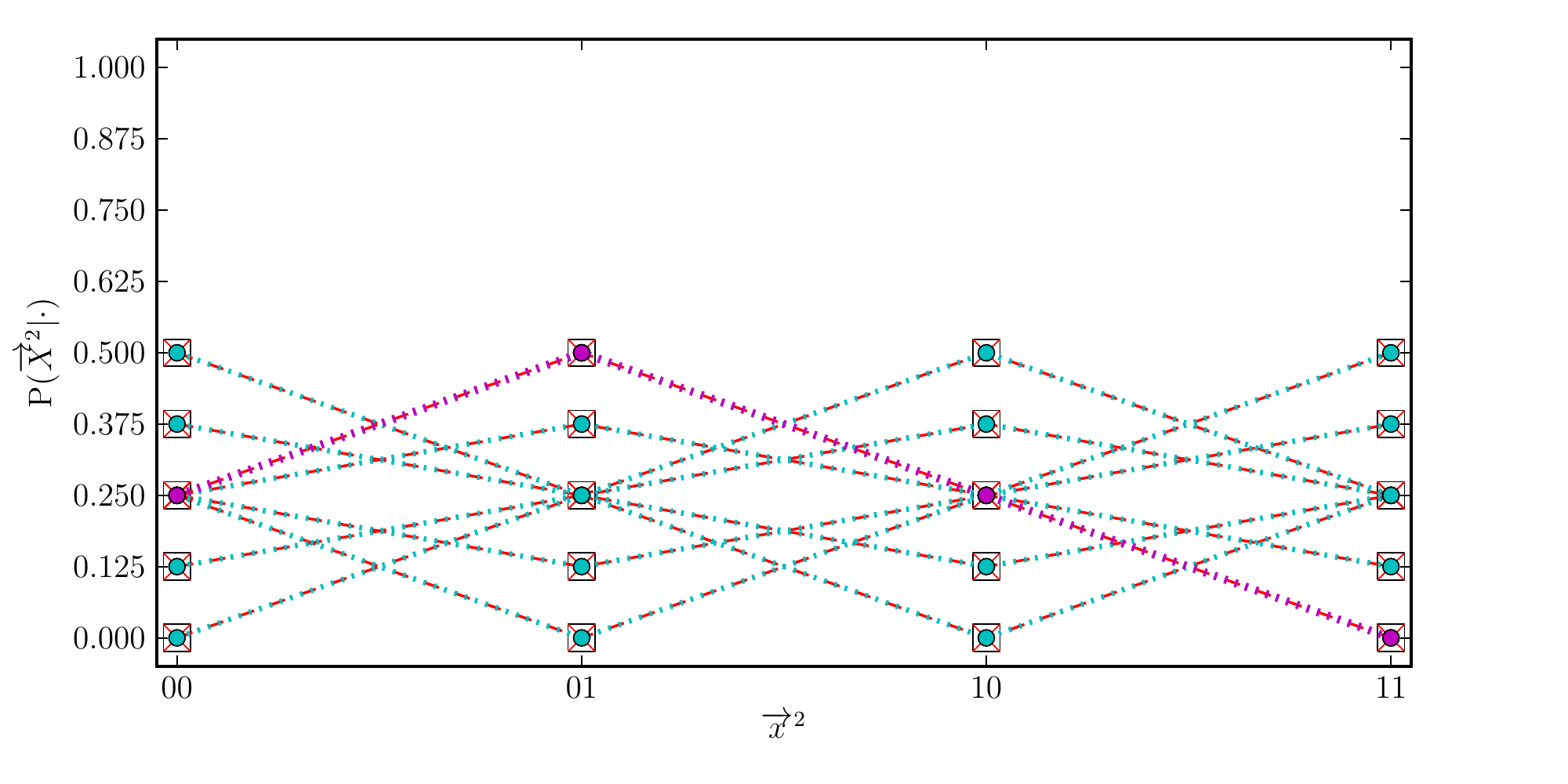}}
\end{center}
\caption{Future conditional probabilities $\Prob(\FinFuture{2}|\cdot)$ for the
  RRXOR process: the $8$-state approximation (circles) finds the causal states
  (boxes). For example, the heavier dashed line (purple) shows
  $\Prob(\FinFuture{2}|\alternatestate) = (1/4,1/2,1/4,0)$.
  Histories of length $K = 3$ were used, along with futures of length $L = 2$.
  }
\label{fig:RRXORMorphsAll}
\end{figure*}

\begin{figure*}
\begin{center}
\resizebox{!}{2.50in}{\includegraphics{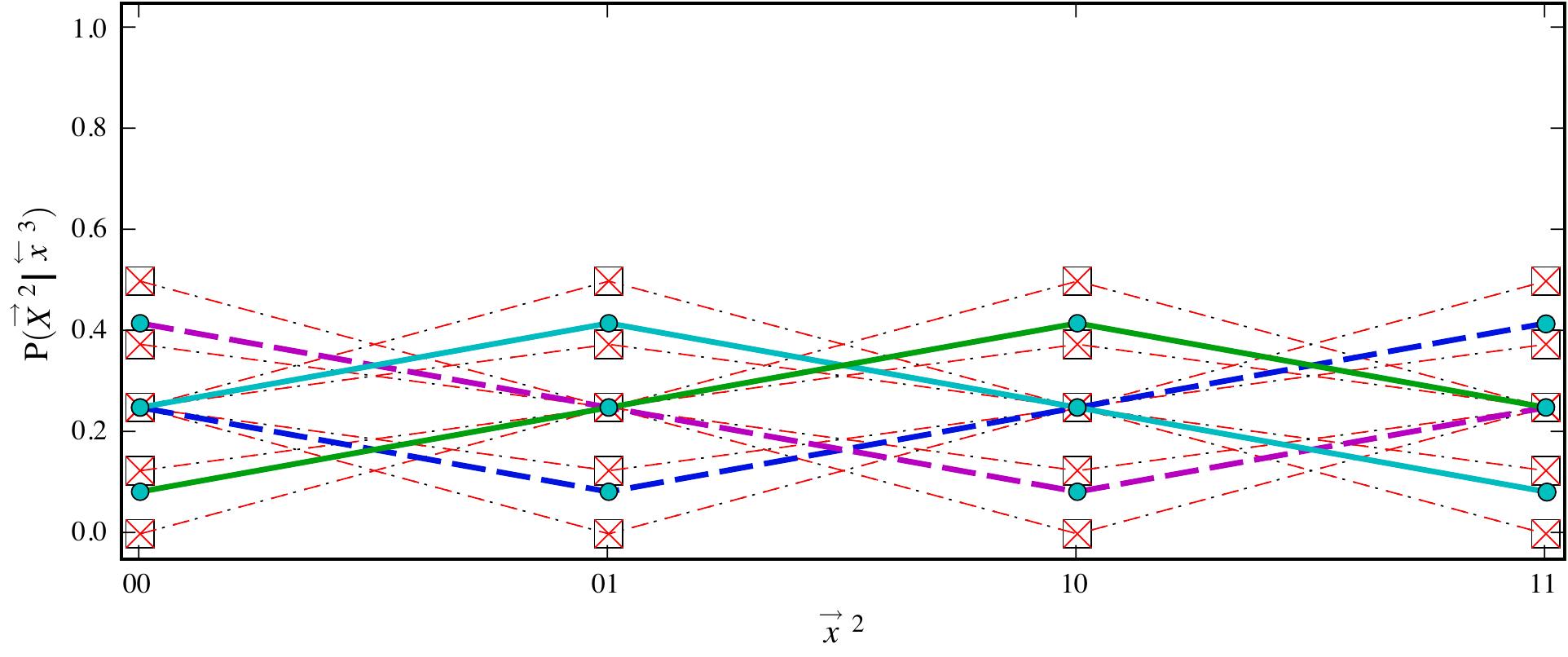}}
\end{center}
\caption{Morphs $\Prob(\FinFuture{2}|\cdot)$ for the RRXOR process: the
  $4$-state approximation (circles and colored lines: state 1 - cyan/full,
  2 - green/full, 3 - blue/dashed, 4 - purple/dashed) compared to causal states
  (boxes). Histories of length $K = 3$ were used, along with
  futures of length $L = 2$.
  }
\label{fig:RRXORMorphs4}
\end{figure*}

As an example, consider the random-random XOR (RRXOR) process which consists of two successive random
symbols chosen to be $0$ or $1$ with equal probability and a third symbol
that is the logical Exclusive-OR (XOR) of the two previous. The RRXOR process
can be represented by a hidden Markov chain with five recurrent causal states,
but having a very large total number of causal states. There are $36$ causal
states, most ($31$) of which describe a complicated transient structure
\citep{Crut01a}. As such, it is a structurally complex process that an
analyst may wish to approximate with a smaller set of states.

Figure \ref{fig:RRXORMInfoPlane} shows the information plane, which specifies
how OCF trades off structure for prediction error as a function of model
complexity for the RRXOR process. The number of effective states (again first
occurrences are denoted by integers along the curve) increases with model
complexity. At a history length of $K = 3$ and future length of $L = 2$, the
process has eight underlying causal states, which are found by
OCF in the $\lambda \rightarrow 0$ limit. The corresponding future conditional
probability distributions are shown in Fig. \ref{fig:RRXORMorphsAll}. 

The RRXOR process has a structure that does not allow for substantial
compression. Fig. \ref{fig:RRXORMInfoPlane} shows that the effective statistical
complexity of the causal-state partition is equal to the full entropy of the
past: $\Cmu (\AlternateState) =  H[\FinPast{3}]$. So, at $L = 3$, unlike the
Even and Golden Mean Processes, the RRXOR process is not compressible.
With half (4) of the number of states, however, OCF reconstructs a
model that is only 33\% as large, while capturing 50\% of the information
about the future. The corresponding conditional future probabilities of the
(best) four-state model are shown in Fig. \ref{fig:RRXORMorphs4}.
They are mixtures of pairs of the eight causal states.

The rate-distortion curve informs the modeler about the (best possible)
efficiency of predictive power to model complexity:
$I[\Partition;\Future] / I[\Past;\Partition]$. This
is useful, for example, if there are constraints on the maximum model size
or, vice versa, on the minimum prediction error. For example, if we require a
model of RRXOR to be 90\% informative about the future, then we can read 
off the curve that this can be achieved at 70\% of the model complexity.
Generally, as $\lambda$ decreases, phase transitions occur to models with a
larger number of effective states \citep{DetermAnneal}. 

\section{Optimal Causal Estimation: Finite-data fluctuations}

In real world applications, we do not know a process's underlying
probability density, but instead must estimate it from a \emph{finite}
time series that we are given. Let that time series be of length $T$ and let us
estimate the joint distribution of pasts (of length $K$) and futures
(of length $L$) via a histogram calculated using a sliding window.
Altogether we have $M = T - ( K + L -1)$ observations.
The resulting estimate $\widehat{\Prob}(\FinPast{K}; \FinFuture{L})$ will
deviate from the true $\Prob(\FinPast{K}; \FinFuture{L})$ by
\mbox{$\Delta(\FinPast{K}, \FinFuture{L})$}. This leads to an overestimate
of the mutual information \footnote{All quantities denoted with a
$\widehat{\cdot}$ are evaluated at the estimate $\widehat{\Prob}$.}:
\mbox{ $\widehat{I}[\FinPast{K};\FinFuture{L}] \geq
I[\FinPast{K};\FinFuture{L}]$}.
Evaluating the objective function at this estimate may lead one to capture
variations that are due to the sampling noise and not to the process's
underlying structure; i.e., OCF may over-fit. That is, the underlying process
may appear to have a larger number $N_c$ of causal states than the true number.

Following Ref. \citep{StillBialek2004}, we argue that this effect can be
counteracted by subtracting from $\widehat{F}[\AlternateState]$ a
model-complexity control term that approximates the error we make by
calculating the estimate $\widehat{F}[\AlternateState]$ rather than the
true $F[\AlternateState]$. If we are willing to assume that $M$ is large
enough, so that the deviation $\Delta(\FinPast{K}, \FinFuture{L})$ is
a small perturbation, then the error can be approximated by
\citep[Eq. (5.8)]{StillBialek2004}:
\begin{equation}
{\cal E} (N_c) = \frac{k^L - 1}{2 \ln(2)} \frac{N_c}{M} ~,
\label{finsizeerror}
\end{equation}
in the low-temperature regime, $\lambda \rightarrow 0$. Recall that $k^L$ is
the total number of possible futures for alphabet size $k$. The optimal number
$N_c^*$ of hidden states is then the one for which the largest amount of mutual
information is shared with the future, corrected by this error:
\begin{equation}
N_c^* :=
  {\rm arg}\max_{N_c} ~ \widehat{I}[\FinPast{K};\FinFuture{L}]_{\lambda \rightarrow 0}^{\rm corrected} (N_c) ~,
\end{equation}
with 
\begin{equation}
\widehat{I}[\FinPast{K};\FinFuture{L}]_{\lambda \rightarrow 0}^{\rm corrected} (N_c)
  = \widehat{I}[\FinPast{K};\FinFuture{L}]_{\lambda \rightarrow 0} (N_c)
  -  {\cal E} (N_c) ~.
\end{equation}
This correction generalizes OCF to \emph{optimal causal estimation} (OCE), a
procedure that simultaneously accounts for the trade-off between structure,
approximation, and sample fluctuations.

\begin{figure*}[ht]
\centering
\resizebox{!}{2.50in}{\includegraphics{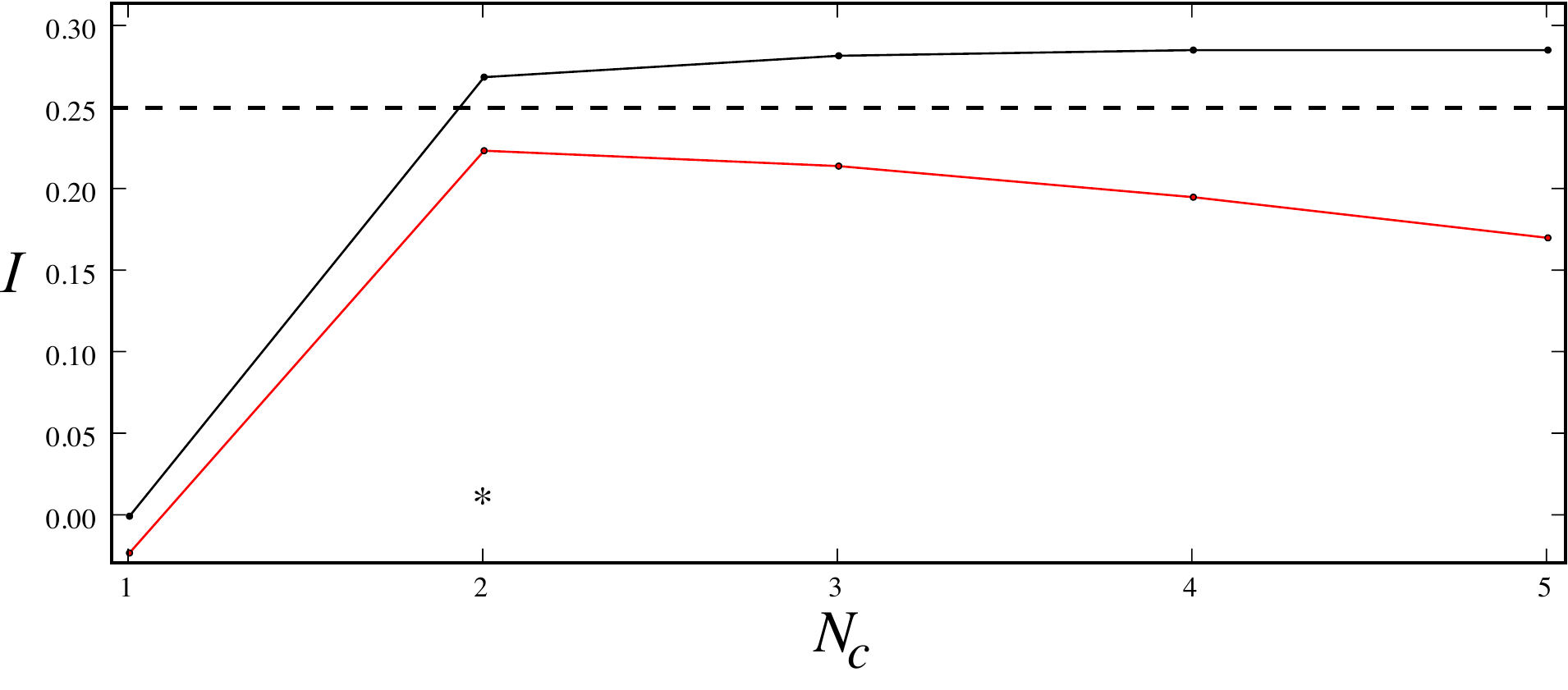}}
\caption{Information $I$ captured about the future versus the number $N_c$ of
  reconstructed states, with statistics estimated from length $T = 100$ time
  series sample from the Golden Mean Process. Upper line: plotted on the
  vertical axis is
  \mbox{$\widehat{I}[\AlternateState;\FinFuture{2}]_{\lambda \rightarrow 0}$}
  (not corrected); lower line: plotted on the vertical axis is the quantity
  \mbox{$\widehat{I}[\AlternateState;\FinFuture{2}]_{\lambda \rightarrow 0}^{\rm corrected}$},
  which is the retained predictive information, but corrected for estimation
  errors due to finite sample size. The dashed line indicates the actual upper
  bound on the predictive information $I[\FinPast{K};\AlternateState]$, for
  comparison. This value is not known to the algorithm, it is computed from
  the true process statistics. Histories of length $K = 3$ and futures of
  length $L = 2$ were used. The asterisk denotes the optimal number
  ($N_c = 2$) of effective states.}
\label{fig:OCEGMPInfoPlane}
\end{figure*}

\begin{figure*}[ht]
\centering
\resizebox{!}{2.50in}{\includegraphics{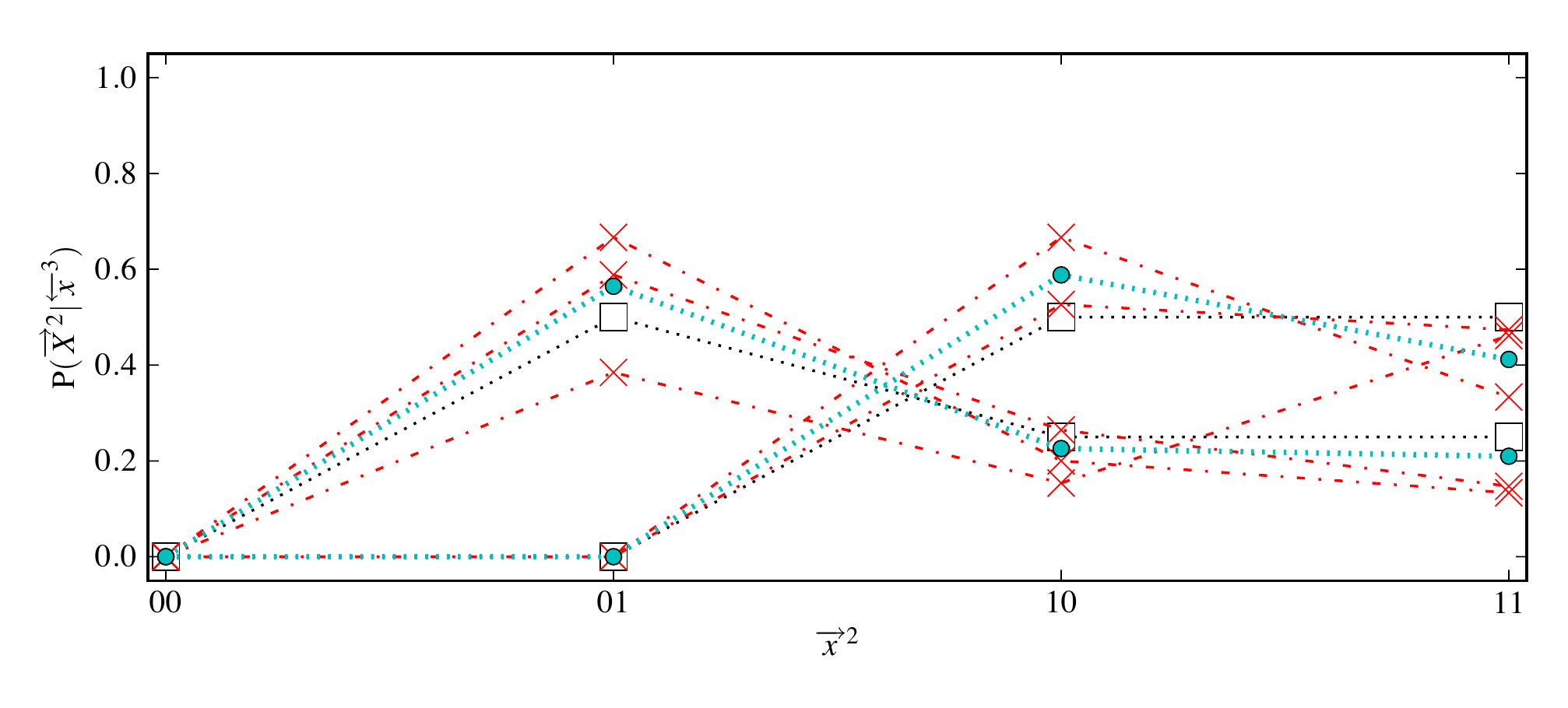}}
\caption{OCE's best two-state approximated future conditional probabilities
  (circles) for the Golden Mean Process. Compared to true (unknown) future
  conditional probabilities (squares). The OCE inputs are the estimates of
  $\widehat{\Prob}(\FinFuture{2}|\finpast{3})$ (crosses).
 }
\label{fig:OCEGMPMorphs}
\end{figure*}

\begin{figure*}[ht]
\centering
\resizebox{!}{2.50in}{\includegraphics{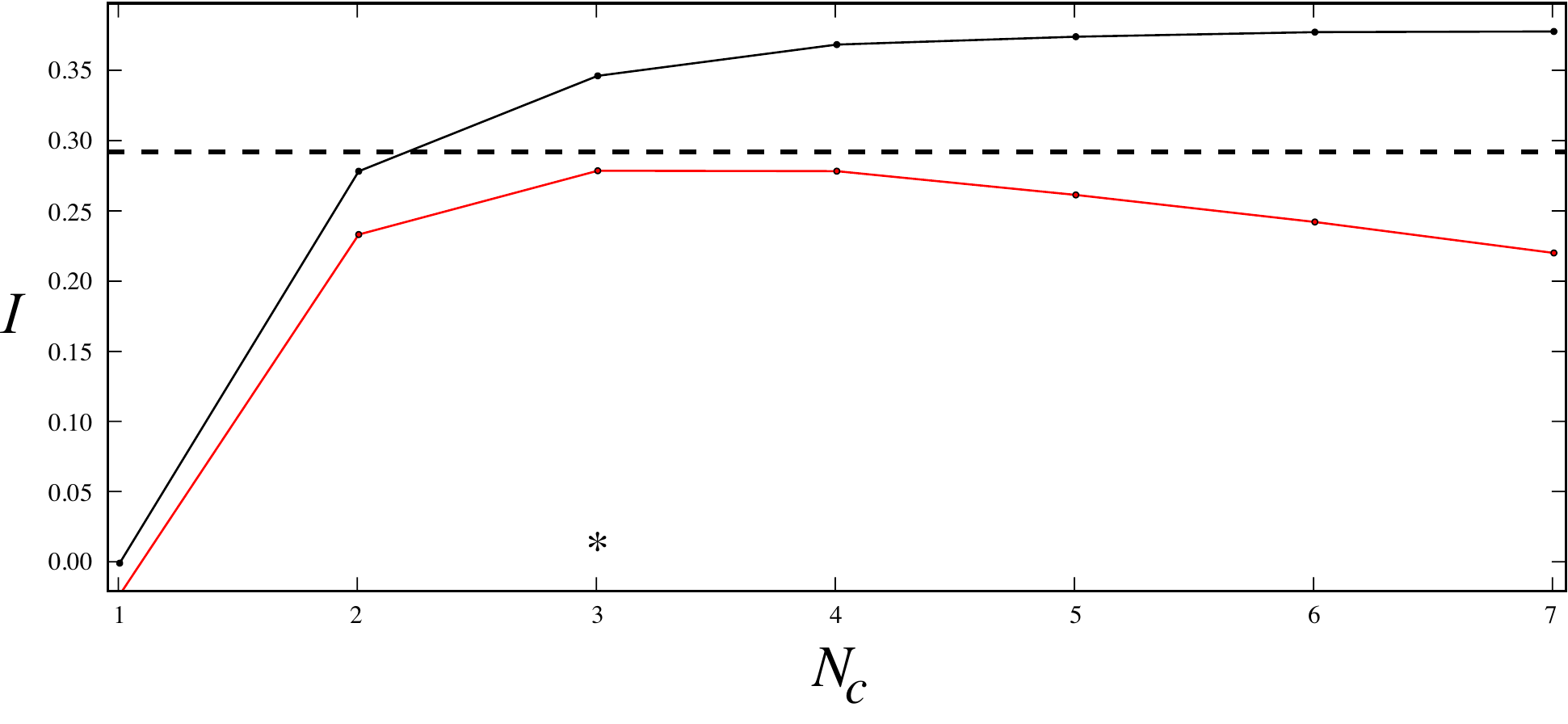}}
\caption{Information $I$ captured about the future versus the number $N_c$ of
  reconstructed states, with statistics estimated from length $T = 100$ time
  series sample from the Even Process. Upper line:
  \mbox{$\widehat{I}[\AlternateState;\FinFuture{2}]_{\lambda \rightarrow 0}$},
  not corrected; lower line: 
  \mbox{$\widehat{I}[\AlternateState;\FinFuture{2}]_{\lambda \rightarrow 0}^{\rm corrected}$},
  corrected for estimation error due to finite sample size. The dashed line
  indicates the actual upper bound on the predictive information, for
  comparison. This value is not known to the algorithm, it is computed from
  the true process statistics. 
  Histories of length $K = 3$ and futures of length $L = 2$ were used.
  The asterisk denotes the optimal number ($N_c = 3$) of effective states.
  }
\label{fig:OCEExampleInfoPlane}
\end{figure*}

\begin{figure*}[ht]
\centering
\resizebox{!}{2.50in}{\includegraphics{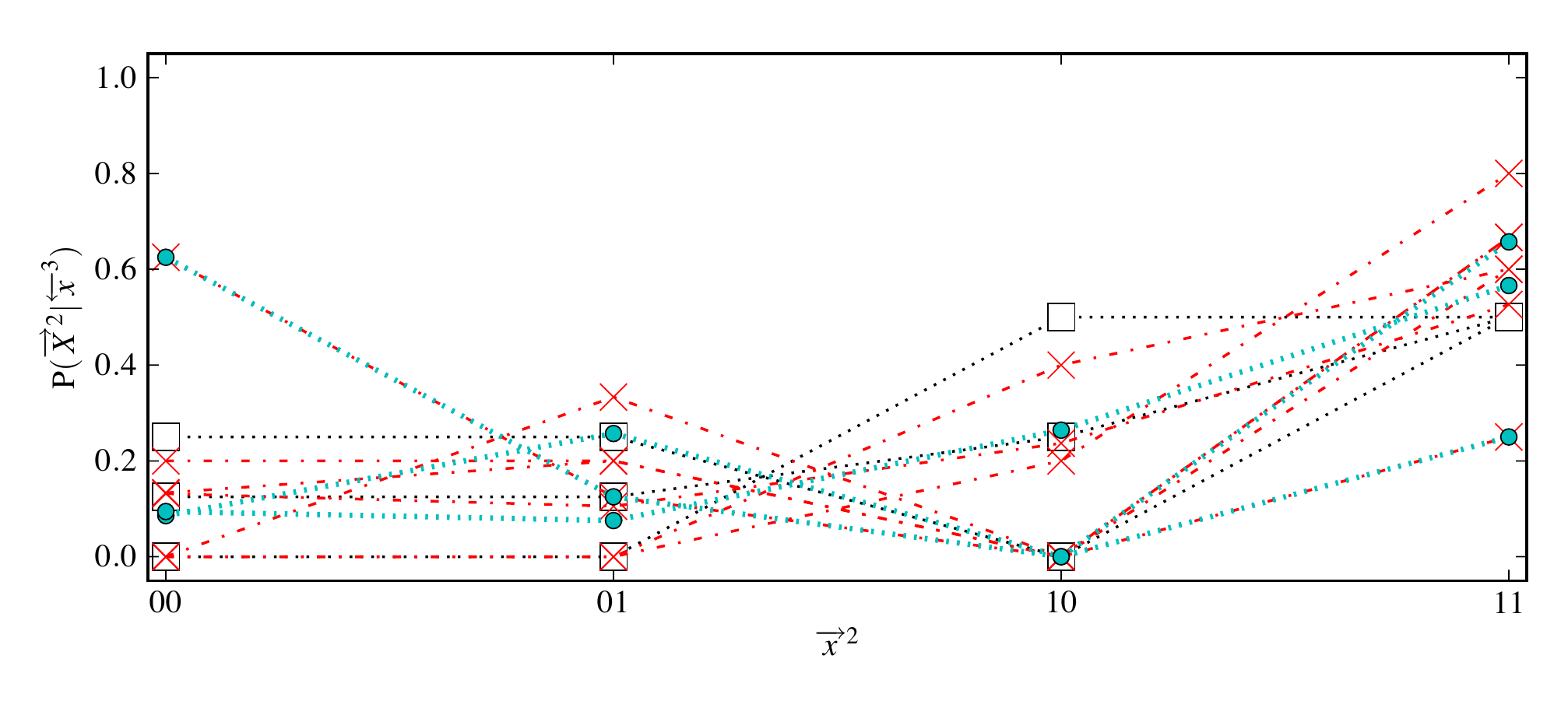}}
\caption{OCE's best three-state approximated future conditional probabilities
  (circles) for the Even Process (d). Compared to true (unknown) future
  conditional probabilities (squares). The OCE inputs are the estimates of
  $\widehat{\Prob}(\FinFuture{2}|\finpast{3})$ (crosses).
 }
\label{fig:OCEExampleMorphs}
\end{figure*}

We illustrate OCE on the Golden Mean and Even Processes studied in Sec.
\ref{examples}. With the {\em correct} number of underlying states, they
can be predicted at a substantial compression. 
Figures \ref{fig:OCEGMPInfoPlane} and \ref{fig:OCEExampleInfoPlane} show
the mutual information $I[\AlternateState;\FinFuture{2}]$ versus the number
$N_c$ of inferred states, with statistics estimated from time series of lengths
$T = 100$. The graphs compare the mutual information
\mbox{$\widehat{I}[\AlternateState;\FinFuture{2}]_{\lambda \rightarrow 0}$}
evaluated using the estimate
\mbox{$\widehat{\Prob}(\FinFuture{2};\FinPast{3})$} (upper curve) to the
corrected information
$\widehat{I}[\AlternateState;\FinFuture{2}]_{\lambda \rightarrow 0}^{\rm corrected}$
calculated by subtracting the approximated error Eq. (\ref{finsizeerror}) with
$k^L = 4$ and $M = 96$ (lower curve).

We see that the corrected information curves peak at, and thereby, select models
with two states for the Golden Mean Process and three states for the Even
Process. This corresponds with the true number of causal states, as we know
from above (Sec. \ref{examples}) for the two processes. The true statistical
complexity for both processes is $\Cmu \approx 0.91830$, while those estimated
via OCE are $\Cmu \approx 0.93773$ and $\Cmu \approx 1.30262$, respectively.
(Recall that the overestimate for the latter was explained in Sec.
\ref{sec:EvenProcess}.)

Figures \ref{fig:OCEGMPMorphs} and \ref{fig:OCEExampleMorphs} show the OCE
future conditional probabilities corresponding to the (optimal) two- and
three-state approximations, respectively. The input to OCE are the future
conditional probabilities given the histories
$\widehat{\Prob}(\FinFuture{2}|\finpast{3})$ (crosses), which are estimated
from the full historical information. Those future conditional probabilities
are corrupted by sampling errors due to the finite data set size and differ
from the true future conditional probabilities (squares). 

Compare the OCE future conditional probabilities (circles) to the true
future conditional probabilities (squares), calculated with the knowledge
of the causal states. (The latter, of course, is not
available to the OCE algorithm.) In the case of the GM Process, OCE
approximates the correct future conditional probabilities. For the Even Process
there is more spread in the estimated OCE future conditional
probabilities. Nonetheless, OCE reduced the fluctuations in its inputs and
corrected in the direction of the true underlying future conditional
probabilities.

\vspace{-.01in}
\section{Conclusion}

We analyzed an information-theoretic approach to causal modeling in two
distinct cases: (i) optimal causal filtering (OCF), where we have access
to the process statistics and desire to capture the process's structure
up to some level of approximation, and (ii) optimal causal estimation
(OCE), in which, in addition, finite-data fluctuations need to be traded-off
against approximation error and structure.
The objective function used in both cases follows from very simple first
principles of information processing and causal modeling: a good model
should minimize prediction error at minimal model complexity. The resulting
principle
of using small, predictive models follows from minimal prior knowledge
that, in particular, makes no structural assumptions about a
process's architecture: Find variables that do the best at causal shielding.

OCF stands in contrast with other approaches. Hidden Markov modeling, for
example, assumes a set of states and an architecture \citep{Rabi86a}. OCF finds
these states from the given data. In minimum description length modeling, to
mention another contrast, the model
complexity of a stochastic source diverges (logarithmically) with the data set
size \citep{Riss89a}, as happens even when modeling the ideal random process
of a fair coin. OCF, however, finds the simplest (smallest) models.

Our main result is that OCF reconstructs the causal-state partition, a
representation previously known from computational mechanics that captures a
process's causal architecture and that allows important system properties,
such as entropy rate and stored information, to be calculated \citep{Crut98d}.
This result is important as it gives a structural meaning to the solutions
of the optimization procedure specified by the causal inference objective
function. We have shown that in the context of time series modeling, where
there is a \emph{natural} relevant variable (the future), the IB approach
\citep{IBN} recovers the unique minimal sufficient statistic---the
causal states---in the limit in
which prediction is paramount to compression. Altogether, this allows us to
go beyond plausibility arguments for the information-theoretic objective
function that have been used. We showed that this way (OCI) of
phrasing the causal inference problem in terms of causal shielding
results in a representation that is a
sufficient statistic and minimal and, moreover, reflects the structure of the
process that generated the data. OCI does so in a way that is meaningful and
well grounded in physics and nonlinear dynamics. The optimal solutions to
balancing prediction and model complexity take on meaning---asymptotically,
they are the causal states. 

The results also contribute to computational mechanics: The continuous trade-off
allows one to extend the deterministic history-to-state assignments that
computational mechanics introduced to ``soft'' partitions of histories. The
theory gives a principled way of constructing stochastic approximations of the
ideal causal architecture. The resulting approximated models can be
substantially smaller and so will be useful in a number of applications.

Finally, we showed how OCF can be adapted to correct for finite-data sampling
fluctuations and so not over-fit. This reduces the tendency to see structure
in noise. OCE finds the correct number of hidden causal states. This renders
the method useful for application to real data.

\section*{Acknowledgments}

UC Davis and the Santa Fe Institute partially supported this work via the
Network Dynamics Program funded by Intel Corporation. It was also partially
supported by the DARPA Physical Intelligence Program. CJE was partially
supported by a Department of Education GAANN graduate fellowship. SS thanks
W. Bialek, discussions with whom have contributed to shaping some of the
ideas expressed, and thanks L. Bottou and I. Nemenmann for useful discussions.


\vspace{-.05in}
\small
\bibliography{chaos,OCIrefs}

\end{document}

%% file: cmechabbrev.tex
\newcommand{\eM}     {$\epsilon$-machine}

\newcommand{\MeasAlphabet}	{\mathcal{A}}
\newcommand{\MeasSymbol}   { {X} }
\newcommand{\meassymbol}   { {x} }
\newcommand{\BiInfinity}	{ \stackrel{\leftrightarrow} {\MeasSymbol} }

\newcommand{\Past}	{ \stackrel{\leftarrow} {\MeasSymbol} }
\newcommand{\past}	{ {\stackrel{\leftarrow} {\meassymbol}} }

\newcommand{\Future}	{ \stackrel{\rightarrow}{\MeasSymbol} }

\newcommand{\AllPasts}	{ { \stackrel{\leftarrow} {\rm {\bf \MeasSymbol}} } }

\newcommand{\CausalState}	{ \mathcal{S} }

\newcommand{\causalstate}	{ \sigma }
\newcommand{\CausalStateSet}	{ \boldsymbol{\CausalState} }
\newcommand{\AlternateState}	{ {\cal R} }
\newcommand{\alternatestate}	{ \rho }
\newcommand{\AlternateStateSet}	{ \boldsymbol{\AlternateState} }
\newcommand{\PrescientState}	{ \widehat{\AlternateState} }

\newcommand{\Prob}		{ {\rm P}}

\newcommand{\Cmu}		{ {C_\mu}}

\newcommand{\EE}		{ {\bf E}}

\newcommand{\InfoGain}[2] { \mathcal{D} \left( {#1} || {#2} \right) }

\newcommand{\ProcessAlphabet}	{\MeasAlphabet}

\newcommand{\FinPast}[1]	{ \stackrel{\leftarrow} {\MeasSymbol} \stackrel{{#1}}{}}
\newcommand{\finpast}[1]  	{ \stackrel{\leftarrow} {\meassymbol}  \stackrel{{#1}}{}}
\newcommand{\FinFuture}[1]		{ \stackrel{\rightarrow}{\MeasSymbol} \stackrel{{#1}}{}}
\newcommand{\finfuture}[1]		{ \stackrel{\rightarrow}{\meassymbol} \stackrel{{#1}}{}}

\newcommand{\Partition}	{ \AlternateState }
\newcommand{\partitionstate}	{ \alternatestate }

\newcommand{\Fdet}   { F_{\rm det} }

%% file: oci.bbl
\begin{thebibliography}{43}
\expandafter\ifx\csname natexlab\endcsname\relax\def\natexlab#1{#1}\fi
\expandafter\ifx\csname bibnamefont\endcsname\relax
  \def\bibnamefont#1{#1}\fi
\expandafter\ifx\csname bibfnamefont\endcsname\relax
  \def\bibfnamefont#1{#1}\fi
\expandafter\ifx\csname citenamefont\endcsname\relax
  \def\citenamefont#1{#1}\fi
\expandafter\ifx\csname url\endcsname\relax
  \def\url#1{\texttt{#1}}\fi
\expandafter\ifx\csname urlprefix\endcsname\relax\def\urlprefix{URL }\fi
\providecommand{\bibinfo}[2]{#2}
\providecommand{\eprint}[2][]{\url{#2}}

\bibitem[{\citenamefont{Berge et~al.}(1986)\citenamefont{Berge, Pomeau, and
  Vidal}}]{Berg84}
\bibinfo{author}{\bibfnamefont{P.}~\bibnamefont{Berge}},
  \bibinfo{author}{\bibfnamefont{Y.}~\bibnamefont{Pomeau}}, \bibnamefont{and}
  \bibinfo{author}{\bibfnamefont{C.}~\bibnamefont{Vidal}},
  \emph{\bibinfo{title}{Order within chaos}} (\bibinfo{publisher}{Wiley},
  \bibinfo{address}{New York}, \bibinfo{year}{1986}).

\bibitem[{\citenamefont{Guckenheimer and Holmes}(1983)}]{Guck83a}
\bibinfo{author}{\bibfnamefont{J.}~\bibnamefont{Guckenheimer}}
  \bibnamefont{and} \bibinfo{author}{\bibfnamefont{P.}~\bibnamefont{Holmes}},
  \emph{\bibinfo{title}{Nonlinear Oscillations, Dynamical Systems, and
  Bifurcations of Vector Fields}} (\bibinfo{publisher}{Springer-Verlag},
  \bibinfo{address}{New York}, \bibinfo{year}{1983}).

\bibitem[{\citenamefont{Wiggins}(1988)}]{Wigg88a}
\bibinfo{author}{\bibfnamefont{S.}~\bibnamefont{Wiggins}},
  \emph{\bibinfo{title}{Global Bifurcations and Chaos: analytical methods}}
  (\bibinfo{publisher}{Springer-Verlag}, \bibinfo{address}{New York},
  \bibinfo{year}{1988}).

\bibitem[{\citenamefont{Devaney}(1989)}]{Deva89a}
\bibinfo{author}{\bibfnamefont{R.~L.} \bibnamefont{Devaney}},
  \emph{\bibinfo{title}{An Introduction to Chaotic Dynamical Systems}}
  (\bibinfo{publisher}{Addison-Wesley}, \bibinfo{address}{Redwood City,
  California}, \bibinfo{year}{1989}).

\bibitem[{\citenamefont{Lieberman and Lichtenberg}(1993)}]{Lieb93a}
\bibinfo{author}{\bibfnamefont{A.~J.} \bibnamefont{Lieberman}}
  \bibnamefont{and} \bibinfo{author}{\bibfnamefont{M.~A.}
  \bibnamefont{Lichtenberg}}, \emph{\bibinfo{title}{Regular and Chaotic
  Dynamics}} (\bibinfo{publisher}{Springer-Verlag}, \bibinfo{address}{New
  York}, \bibinfo{year}{1993}), \bibinfo{edition}{2nd} ed.

\bibitem[{\citenamefont{Ott}(1993)}]{Ott93a}
\bibinfo{author}{\bibfnamefont{E.}~\bibnamefont{Ott}},
  \emph{\bibinfo{title}{Chaos in Dynamical Systems}}
  (\bibinfo{publisher}{Cambridge University Press}, \bibinfo{address}{New
  York}, \bibinfo{year}{1993}).

\bibitem[{\citenamefont{Strogatz}(1994)}]{Stro94a}
\bibinfo{author}{\bibfnamefont{S.~H.} \bibnamefont{Strogatz}},
  \emph{\bibinfo{title}{Nonlinear Dynamics and Chaos: with applications to
  physics, biology, chemistry, and engineering}}
  (\bibinfo{publisher}{Addison-Wesley}, \bibinfo{address}{Reading,
  Massachusetts}, \bibinfo{year}{1994}).

\bibitem[{\citenamefont{Packard et~al.}(1980)\citenamefont{Packard,
  Crutchfield, Farmer, and Shaw}}]{Pack80}
\bibinfo{author}{\bibfnamefont{N.~H.} \bibnamefont{Packard}},
  \bibinfo{author}{\bibfnamefont{J.~P.} \bibnamefont{Crutchfield}},
  \bibinfo{author}{\bibfnamefont{J.~D.} \bibnamefont{Farmer}},
  \bibnamefont{and} \bibinfo{author}{\bibfnamefont{R.~S.} \bibnamefont{Shaw}},
  \bibinfo{journal}{Phys. Rev. Let.} \textbf{\bibinfo{volume}{45}},
  \bibinfo{pages}{712} (\bibinfo{year}{1980}).

\bibitem[{\citenamefont{Takens}(1981)}]{Take81}
\bibinfo{author}{\bibfnamefont{F.}~\bibnamefont{Takens}}, in
  \emph{\bibinfo{booktitle}{Symposium on Dynamical Systems and Turbulence}},
  edited by \bibinfo{editor}{\bibfnamefont{D.~A.} \bibnamefont{Rand}}
  \bibnamefont{and} \bibinfo{editor}{\bibfnamefont{L.~S.} \bibnamefont{Young}}
  (\bibinfo{publisher}{Springer-Verlag}, \bibinfo{address}{Berlin},
  \bibinfo{year}{1981}), vol. \bibinfo{volume}{898}, p. \bibinfo{pages}{366}.

\bibitem[{\citenamefont{Fraser}(1991)}]{Fras90b}
\bibinfo{author}{\bibfnamefont{A.}~\bibnamefont{Fraser}}, in
  \emph{\bibinfo{booktitle}{Information Dynamics}}, edited by
  \bibinfo{editor}{\bibfnamefont{H.}~\bibnamefont{Atmanspacher}}
  \bibnamefont{and}
  \bibinfo{editor}{\bibfnamefont{H.}~\bibnamefont{Scheingraber}}
  (\bibinfo{publisher}{Plenum}, \bibinfo{address}{New York},
  \bibinfo{year}{1991}), vol. \bibinfo{volume}{Series B: Physics Vol. 256} of
  \emph{\bibinfo{series}{NATO ASI Series}}, p. \bibinfo{pages}{125}.

\bibitem[{\citenamefont{Crutchfield and McNamara}(1987)}]{Crut87a}
\bibinfo{author}{\bibfnamefont{J.~P.} \bibnamefont{Crutchfield}}
  \bibnamefont{and} \bibinfo{author}{\bibfnamefont{B.~S.}
  \bibnamefont{McNamara}}, \bibinfo{journal}{Complex Systems}
  \textbf{\bibinfo{volume}{1}}, \bibinfo{pages}{417 } (\bibinfo{year}{1987}).

\bibitem[{\citenamefont{Casdagli and Eubank}(1992)}]{Casd91a}
\bibinfo{editor}{\bibfnamefont{M.}~\bibnamefont{Casdagli}} \bibnamefont{and}
  \bibinfo{editor}{\bibfnamefont{S.}~\bibnamefont{Eubank}}, eds.,
  \emph{\bibinfo{title}{Nonlinear Modeling}}, SFI Studies in the Sciences of
  Complexity (\bibinfo{publisher}{Addison-Wesley}, \bibinfo{address}{Reading,
  Massachusetts}, \bibinfo{year}{1992}).

\bibitem[{\citenamefont{Sprott}(2003)}]{Spro03a}
\bibinfo{author}{\bibfnamefont{J.~C.} \bibnamefont{Sprott}},
  \emph{\bibinfo{title}{Chaos and Time-Series Analysis}}
  (\bibinfo{publisher}{Oxford University Press}, \bibinfo{address}{Oxford, UK},
  \bibinfo{year}{2003}), \bibinfo{edition}{2nd} ed.

\bibitem[{\citenamefont{Kantz and Schreiber}(2006)}]{Kant06a}
\bibinfo{author}{\bibfnamefont{H.}~\bibnamefont{Kantz}} \bibnamefont{and}
  \bibinfo{author}{\bibfnamefont{T.}~\bibnamefont{Schreiber}},
  \emph{\bibinfo{title}{Nonlinear Time Series Analysis}}
  (\bibinfo{publisher}{Cambridge University Press},
  \bibinfo{address}{Cambridge, UK}, \bibinfo{year}{2006}),
  \bibinfo{edition}{2nd} ed.

\bibitem[{\citenamefont{Crutchfield and Young}(1989)}]{Crut88a}
\bibinfo{author}{\bibfnamefont{J.~P.} \bibnamefont{Crutchfield}}
  \bibnamefont{and} \bibinfo{author}{\bibfnamefont{K.}~\bibnamefont{Young}},
  \bibinfo{journal}{Phys. Rev. Let.} \textbf{\bibinfo{volume}{63}},
  \bibinfo{pages}{105} (\bibinfo{year}{1989}).

\bibitem[{\citenamefont{Tishby et~al.}(1999)\citenamefont{Tishby, Pereira, and
  Bialek}}]{IBN}
\bibinfo{author}{\bibfnamefont{N.}~\bibnamefont{Tishby}},
  \bibinfo{author}{\bibfnamefont{F.}~\bibnamefont{Pereira}}, \bibnamefont{and}
  \bibinfo{author}{\bibfnamefont{W.}~\bibnamefont{Bialek}}, in
  \emph{\bibinfo{booktitle}{Proceedings of the 37th Annual Allerton
  Conference}}, edited by
  \bibinfo{editor}{\bibfnamefont{B.}~\bibnamefont{Hajek}} \bibnamefont{and}
  \bibinfo{editor}{\bibfnamefont{R.~S.} \bibnamefont{Sreenivas}}
  (\bibinfo{publisher}{University of Illinois}, \bibinfo{year}{1999}), pp.
  \bibinfo{pages}{368--377}.

\bibitem[{\citenamefont{Still and Bialek}(2004)}]{StillBialek2004}
\bibinfo{author}{\bibfnamefont{S.}~\bibnamefont{Still}} \bibnamefont{and}
  \bibinfo{author}{\bibfnamefont{W.}~\bibnamefont{Bialek}},
  \bibinfo{journal}{Neural Computation} \textbf{\bibinfo{volume}{16(12)}},
  \bibinfo{pages}{2483} (\bibinfo{year}{2004}).

\bibitem[{\citenamefont{Shalizi and Crutchfield}(2002)}]{Shal99a}
\bibinfo{author}{\bibfnamefont{C.~R.} \bibnamefont{Shalizi}} \bibnamefont{and}
  \bibinfo{author}{\bibfnamefont{J.~P.} \bibnamefont{Crutchfield}},
  \bibinfo{journal}{Advances in Complex Systems} \textbf{\bibinfo{volume}{5}},
  \bibinfo{pages}{1} (\bibinfo{year}{2002}).

\bibitem[{\citenamefont{Shannon}(1948)}]{Shannon48}
\bibinfo{author}{\bibfnamefont{C.~E.} \bibnamefont{Shannon}},
  \bibinfo{journal}{Bell Sys. Tech. J.} \textbf{\bibinfo{volume}{27}}
  (\bibinfo{year}{1948}), \bibinfo{note}{reprinted in C. E. Shannon and W.
  Weaver {\it The Mathematical Theory of Communication}, University of Illinois
  Press, Urbana, 1949}.

\bibitem[{\citenamefont{Still and Crutchfield}(2007)}]{Still07a}
\bibinfo{author}{\bibfnamefont{S.}~\bibnamefont{Still}} \bibnamefont{and}
  \bibinfo{author}{\bibfnamefont{J.~P.} \bibnamefont{Crutchfield}}
  (\bibinfo{year}{2007}), \bibinfo{note}{arxiv.org: 0708.0654
  [physics.gen-ph]}.

\bibitem[{\citenamefont{Crutchfield}(1994)}]{Crut92c}
\bibinfo{author}{\bibfnamefont{J.~P.} \bibnamefont{Crutchfield}},
  \bibinfo{journal}{Physica D} \textbf{\bibinfo{volume}{75}},
  \bibinfo{pages}{11} (\bibinfo{year}{1994}).

\bibitem[{\citenamefont{Crutchfield and Shalizi}(1999)}]{Crut98d}
\bibinfo{author}{\bibfnamefont{J.~P.} \bibnamefont{Crutchfield}}
  \bibnamefont{and} \bibinfo{author}{\bibfnamefont{C.~R.}
  \bibnamefont{Shalizi}}, \bibinfo{journal}{Physical Review E}
  \textbf{\bibinfo{volume}{59}}, \bibinfo{pages}{275} (\bibinfo{year}{1999}).

\bibitem[{\citenamefont{Cover and Thomas}(2006)}]{Cove06a}
\bibinfo{author}{\bibfnamefont{T.~M.} \bibnamefont{Cover}} \bibnamefont{and}
  \bibinfo{author}{\bibfnamefont{J.~A.} \bibnamefont{Thomas}},
  \emph{\bibinfo{title}{Elements of Information Theory}}
  (\bibinfo{publisher}{Wiley-Interscience}, \bibinfo{address}{New York},
  \bibinfo{year}{2006}), \bibinfo{edition}{2nd} ed.

\bibitem[{\citenamefont{Crutchfield et~al.}(2009)\citenamefont{Crutchfield,
  Ellison, and Mahoney}}]{Crut08a}
\bibinfo{author}{\bibfnamefont{J.~P.} \bibnamefont{Crutchfield}},
  \bibinfo{author}{\bibfnamefont{C.~J.} \bibnamefont{Ellison}},
  \bibnamefont{and} \bibinfo{author}{\bibfnamefont{J.~R.}
  \bibnamefont{Mahoney}}, \bibinfo{journal}{Phys. Rev. Lett.}
  \textbf{\bibinfo{volume}{103}}, \bibinfo{pages}{094101}
  (\bibinfo{year}{2009}).

\bibitem[{\citenamefont{Crutchfield and Feldman}(2003)}]{Crut01a}
\bibinfo{author}{\bibfnamefont{J.~P.} \bibnamefont{Crutchfield}}
  \bibnamefont{and} \bibinfo{author}{\bibfnamefont{D.~P.}
  \bibnamefont{Feldman}}, \bibinfo{journal}{CHAOS}
  \textbf{\bibinfo{volume}{13}}, \bibinfo{pages}{25} (\bibinfo{year}{2003}).

\bibitem[{\citenamefont{Bialek et~al.}(2006)\citenamefont{Bialek, de~Ruyter~van
  Steveninck, and Tishby}}]{bialek06}
\bibinfo{author}{\bibfnamefont{W.}~\bibnamefont{Bialek}},
  \bibinfo{author}{\bibfnamefont{R.~R.} \bibnamefont{de~Ruyter~van
  Steveninck}}, \bibnamefont{and}
  \bibinfo{author}{\bibfnamefont{N.}~\bibnamefont{Tishby}}, in
  \emph{\bibinfo{booktitle}{Proceedings of the International Symposium on
  Information Theory}} (\bibinfo{year}{2006}), pp. \bibinfo{pages}{659--663}.

\bibitem[{\citenamefont{del Junco and Rahe}(1979)}]{Junc79}
\bibinfo{author}{\bibfnamefont{A.}~\bibnamefont{del Junco}} \bibnamefont{and}
  \bibinfo{author}{\bibfnamefont{M.}~\bibnamefont{Rahe}},
  \bibinfo{journal}{Proc. AMS} \textbf{\bibinfo{volume}{75}},
  \bibinfo{pages}{259} (\bibinfo{year}{1979}).

\bibitem[{\citenamefont{Crutchfield and Packard}(1983)}]{Crut83a}
\bibinfo{author}{\bibfnamefont{J.~P.} \bibnamefont{Crutchfield}}
  \bibnamefont{and} \bibinfo{author}{\bibfnamefont{N.~H.}
  \bibnamefont{Packard}}, \bibinfo{journal}{Physica}
  \textbf{\bibinfo{volume}{7D}}, \bibinfo{pages}{201 } (\bibinfo{year}{1983}).

\bibitem[{\citenamefont{Shaw}(1984)}]{Shaw84}
\bibinfo{author}{\bibfnamefont{R.}~\bibnamefont{Shaw}},
  \emph{\bibinfo{title}{The Dripping Faucet as a Model Chaotic System}}
  (\bibinfo{publisher}{Aerial Press}, \bibinfo{address}{Santa Cruz,
  California}, \bibinfo{year}{1984}).

\bibitem[{\citenamefont{Grassberger}(1986)}]{Gras86}
\bibinfo{author}{\bibfnamefont{P.}~\bibnamefont{Grassberger}},
  \bibinfo{journal}{Intl. J. Theo. Phys.} \textbf{\bibinfo{volume}{25}},
  \bibinfo{pages}{907} (\bibinfo{year}{1986}).

\bibitem[{\citenamefont{Li}(1991)}]{Li91}
\bibinfo{author}{\bibfnamefont{W.}~\bibnamefont{Li}}, \bibinfo{journal}{Complex
  Systems} \textbf{\bibinfo{volume}{5}}, \bibinfo{pages}{381}
  (\bibinfo{year}{1991}).

\bibitem[{\citenamefont{Bialek and Tishby}(1999)}]{BT99}
\bibinfo{author}{\bibfnamefont{W.}~\bibnamefont{Bialek}} \bibnamefont{and}
  \bibinfo{author}{\bibfnamefont{N.}~\bibnamefont{Tishby}},
  \emph{\bibinfo{title}{Predictive information}} (\bibinfo{year}{1999}),
  \urlprefix\url{arXiv:cond-mat/9902341v1}.

\bibitem[{\citenamefont{Crutchfield et~al.}(2010)\citenamefont{Crutchfield,
  Ellison, Mahoney, and James}}]{Crut10a}
\bibinfo{author}{\bibfnamefont{J.~P.} \bibnamefont{Crutchfield}},
  \bibinfo{author}{\bibfnamefont{C.~J.} \bibnamefont{Ellison}},
  \bibinfo{author}{\bibfnamefont{J.~R.} \bibnamefont{Mahoney}},
  \bibnamefont{and} \bibinfo{author}{\bibfnamefont{R.~G.} \bibnamefont{James}},
  \bibinfo{journal}{CHAOS} p. \bibinfo{pages}{in press} (\bibinfo{year}{2010}).

\bibitem[{\citenamefont{Arimoto}(1972)}]{Arimoto72}
\bibinfo{author}{\bibfnamefont{S.}~\bibnamefont{Arimoto}},
  \bibinfo{journal}{IEEE Transactions on Information Theory IT-18} pp.
  \bibinfo{pages}{14--20} (\bibinfo{year}{1972}).

\bibitem[{\citenamefont{Blahut}(1972)}]{Blahut72}
\bibinfo{author}{\bibfnamefont{R.~E.} \bibnamefont{Blahut}},
  \bibinfo{journal}{IEEE Transactions on Information Theory IT-18} pp.
  \bibinfo{pages}{460--473} (\bibinfo{year}{1972}).

\bibitem[{\citenamefont{Rose et~al.}(1990)\citenamefont{Rose, Gurewitz, and
  Fox}}]{Rose90}
\bibinfo{author}{\bibfnamefont{K.}~\bibnamefont{Rose}},
  \bibinfo{author}{\bibfnamefont{E.}~\bibnamefont{Gurewitz}}, \bibnamefont{and}
  \bibinfo{author}{\bibfnamefont{G.~C.} \bibnamefont{Fox}},
  \bibinfo{journal}{Phys. Rev. Lett.} \textbf{\bibinfo{volume}{65}},
  \bibinfo{pages}{945} (\bibinfo{year}{1990}).

\bibitem[{\citenamefont{Rose}(1998)}]{DetermAnneal}
\bibinfo{author}{\bibfnamefont{K.}~\bibnamefont{Rose}}, \bibinfo{journal}{Proc.
  of the IEEE} \textbf{\bibinfo{volume}{86}}, \bibinfo{pages}{2210}
  (\bibinfo{year}{1998}).

\bibitem[{\citenamefont{Ellison et~al.}(2009)\citenamefont{Ellison, Mahoney,
  and Crutchfield}}]{Crut08b}
\bibinfo{author}{\bibfnamefont{C.~J.} \bibnamefont{Ellison}},
  \bibinfo{author}{\bibfnamefont{J.~R.} \bibnamefont{Mahoney}},
  \bibnamefont{and} \bibinfo{author}{\bibfnamefont{J.~P.}
  \bibnamefont{Crutchfield}}, \bibinfo{journal}{J. Stat. Phys.}
  \textbf{\bibinfo{volume}{136}}, \bibinfo{pages}{1005} (\bibinfo{year}{2009}).

\bibitem[{\citenamefont{Strelioff et~al.}(2007)\citenamefont{Strelioff,
  Crutchfield, and H{\"u}bler}}]{Stre07a}
\bibinfo{author}{\bibfnamefont{C.~C.} \bibnamefont{Strelioff}},
  \bibinfo{author}{\bibfnamefont{J.~P.} \bibnamefont{Crutchfield}},
  \bibnamefont{and}
  \bibinfo{author}{\bibfnamefont{A.}~\bibnamefont{H{\"u}bler}},
  \bibinfo{journal}{Phys. Rev. E} \textbf{\bibinfo{volume}{76}},
  \bibinfo{pages}{011106} (\bibinfo{year}{2007}).

\bibitem[{\citenamefont{Rabiner and Juang}(1986)}]{Rabi86a}
\bibinfo{author}{\bibfnamefont{L.~R.} \bibnamefont{Rabiner}} \bibnamefont{and}
  \bibinfo{author}{\bibfnamefont{B.~H.} \bibnamefont{Juang}},
  \bibinfo{journal}{IEEE ASSP Magazine} \textbf{\bibinfo{volume}{January}}
  (\bibinfo{year}{1986}).

\bibitem[{\citenamefont{Rissanen}(1989)}]{Riss89a}
\bibinfo{author}{\bibfnamefont{J.}~\bibnamefont{Rissanen}},
  \emph{\bibinfo{title}{Stochastic Complexity in Statistical Inquiry}}
  (\bibinfo{publisher}{World Scientific}, \bibinfo{address}{Singapore},
  \bibinfo{year}{1989}).

\bibitem[{\citenamefont{Still}(2009)}]{Still09IAL}
\bibinfo{author}{\bibfnamefont{S.}~\bibnamefont{Still}},
  \bibinfo{journal}{Euro. Phys. Lett.} \textbf{\bibinfo{volume}{85}},
  \bibinfo{pages}{28005} (\bibinfo{year}{2009}).

\bibitem[{\citenamefont{Ay and Crutchfield}(2005)}]{Ay05a}
\bibinfo{author}{\bibfnamefont{N.}~\bibnamefont{Ay}} \bibnamefont{and}
  \bibinfo{author}{\bibfnamefont{J.~P.} \bibnamefont{Crutchfield}},
  \bibinfo{journal}{J. Stat. Phys.} \textbf{\bibinfo{volume}{210}},
  \bibinfo{pages}{659} (\bibinfo{year}{2005}).

\end{thebibliography}
